\documentclass[aps,prl,reprint,twocolumn,superscriptaddress,showpacs,longbibliography]{revtex4-2}
\usepackage{graphicx,amsbsy,amsmath,epsfig,natbib,float}
\usepackage{amssymb,latexsym,wasysym,pifont,mathrsfs}
\usepackage{comment,verbatim,multirow,array,sidecap,color}
\usepackage[normalem]{ulem}

\definecolor{blueberry}{rgb}{0.01569, 0.2, 1.0}
\definecolor{asparagus}{rgb}{0.4,0.4,0}
\definecolor{maraschino}{rgb}{1.0,0.0,0.0}
\definecolor{strawberry}{rgb}{1.0,0.0,0.5}

\definecolor{mygreen}{rgb}{0.0,0.55,0.3}

\begin{document}

\title{Yielding in colloidal gels: from local structure to meso-scale strand breakage and macroscopic failure}

\author{Himangsu Bhaumik}
\affiliation{Yusuf Hamied Department of Chemistry, University of Cambridge, Lensfield Road, Cambridge CB2 1EW, UK}

\author{Tanniemola B. Liverpool}
\affiliation{School of Mathematics, University of Bristol, Fry Building, Bristol BS8 1UG, UK}

\author{C. Patrick Royall}
\affiliation{H.H. Wills Physics Laboratory, Tyndall Avenue, Bristol, BS8 1TL, UK}
\affiliation{School of Chemistry, University of Bristol, Cantock’s Close, Bristol, BS8 1TS, UK}
\affiliation{Gulliver UMR CNRS 7083, ESPCI Paris, Université PSL, 75005 Paris, France}

\author{Robert L. Jack}
\affiliation{Yusuf Hamied Department of Chemistry, University of Cambridge, Lensfield Road, Cambridge CB2 1EW, UK}
\affiliation{DAMTP, Centre for Mathematical Sciences, University of Cambridge, Wilberforce Road, Cambridge CB3 0WA, UK}

\date{\today}
 
\begin{abstract}  
We study creep flow and yielding of particulate depletion gels under constant shear stress, combining data on different length and time scales.   We characterise the breakage of meso-scale strands in the gel.  Breakage events are distributed homogeneously in space, corresponding to macroscopically ductile flow.  At the microscale, a spatio-temporal analysis of structural and mechanical metrics connects properties of strands before and after they fail, indicating that strand breakage is statistically predictable. Using results from different scales, we discuss the interplay between creeping and aging dynamics, and we demonstrate a viscosity bifurcation.
\end{abstract}

\maketitle

\newcommand{\SM}{Appendices}
\newcommand{\SMref}{Appendices}
\newcommand{\SMlong}{Appendices}

\newcommand{\eps}{\varepsilon}
\newcommand{\eprep}{\eps_{0,\rm prep}}

Colloidal particles with strong attractive interactions can undergo gelation, forming kinetically-arrested networks of strands~\cite{zaccarelli2007colloidal,
royall2021real,
cipelletti2005slow,
poon2002physics,
lu2008gelation}.
These materials have widespread industrial applications, for example in food \cite{mezzenga2005understanding}, paints \cite{Xiong2019}, and biomedical engineering \cite{diba2017highly,rose2014}.  
Gels are non-equilibrium systems whose properties depend strongly on their preparation and their history.  
Their rheology exhibits complex features typically seen in soft solids~\cite{BonnRMP2017,DivouxetalSMperspective24,berthier2024yielding,cochran2024}, and is also coupled to their characteristic coarsening and aging properties~\cite{BonnScience09,bouzid2017elastically,aime2018microscopic,lockwood2024}.
Such effects mean that formulation of gel products is difficult to predict and control.
For example, the delayed gravitational collapse of some gels poses important challenges for product shelf life~\cite{Starrs2002,bartlett2012sudden,Varga2018}.
 
From a theoretical and computational perspective, prediction of gels' properties is challenging because of a range of relevant length scales -- macroscopic rheology and gel collapse depend on the (mesoscopic) gel strands, and these depend in turn on the individual particle interactions, which are the microscopic control parameters.  
An overarching theoretical challenge is to bridge scales from gels' microscopic structure to their rheology, presumably with mesoscale strands' behaviour as an intermediate step.
Important questions within this area include: identification of relevant topological and mechanical features of the network of strands~\cite{WeitzPRL06, HsiaoPNAS12, bouzid2017elastically, tsurusawa2019direct,bantawa2022hidden,JamaliPNAS24,JamaliPRE24}; characterisation of strand breakage~\cite{van2018strand,verweij2019plasticity,KrisSMNeck23} and its relation to macroscopic yielding~\cite{sprakel2011stress,lindstrom2012structures,colombo2014stress,landrum2016delayed,SaintSM17ProteinGel}; understanding structural properties of strands
and their dependence on microscopic particle interactions~\cite{lu2008gelation,patrick2008direct,griffiths2017local,MullerNatCom2023,Mangal2024}; and connecting aging and sample history to its rheology~\cite{BonnPRL2002,Coussot2002,BonnScience09,zia2014micro,KoumakisSM15}.

This work analyses depletion gels~\cite{poon2002physics,lu2008gelation}, where bond formation is reversible, due to thermal fluctuations.  These may be contrasted with irreversible gels, as formed (for example) through van der Waals interactions~\cite{WeitzPRL2000Universal},  leading to fractal structures whose strand thicknesses {may be}
only one or two particles, as also found in computational model gels with directional {interactions}~\cite{colombo2014stress,bouzid2017elastically,bantawa2022hidden}.  In depletion gels, which are formed by arrested spinodal decomposition, one finds thicker strands, whose internal structure is similar to colloidal glasses~\cite{zaccarelli2007colloidal,patrick2008direct,royall2021real,Fenton2023}.

The perspective of~\cite{sprakel2011stress,lindstrom2012structures} is that yielding of fractal gels takes place by stretching and breakage of strands. Individual breakage events have been simulated numerically~\cite{van2018strand,verweij2019plasticity,KrisSMNeck23}, and observed in particle-resolved experiments~\cite{tsurusawa2019direct}. 
Simulations of yielding in a model fractal gel~\cite{bouzid2017elastically} show similar events.  However, reversible depletion gels are more complicated: particle motion within the strands can enable yielding with very few breakages~\cite{landrum2016delayed}, but this depends on strands' local structure and bonding. 

In this work, we consider
 an accurate simulation model of a depletion gel under applied shear stress.  
We combine results on different length scales, to analyse yielding and failure.
We characterise strand-breaking events using a statistical mechanical framework, based on extensive numerical data.  We find  distinctive structural signatures of strand breakage which manifest significantly before yielding, offering new possibilities for prediction of gel behaviour.  
Strand-breaking events are distributed homogeneously in space, indicating that the macroscopic yielding process has a ductile character, in contrast to the brittle behaviour of fractal (irreversible) gels~\cite{colombo2014stress,LeomachPRL14,SaintSM17ProteinGel}.
By considering the interplay of aging and macroscopic yielding, we also demonstrate a viscosity bifurcation in numerical simulations, complementing the experimental and theoretical perspectives of~\cite{BonnPRL2002,Coussot2002,BonnScience09}.
 
Together, these results -- especially the systematic characterisation of strand-breaking -- represent an important step in bridging scales between microscopic local structure, mesoscopic strand-breaking events, and macroscopic rheology.  We discuss how they pave the way for a unified theory of creeping and yielding in reversible gels, for example via coarse-grained mesoscopic models.

\emph{Model}.  We simulate a three-dimensional size-polydisperse colloid-polymer mixture in a periodic box of volume $V$, which accurately mimics experimental gel-formers~\cite{taffs2010,royall2018vitrification}. 
{Effective interactions between colloidal particles are described by a} non-dimensionalized Morse potential
\begin{equation}
U_0(r)=\varepsilon_0\big[e^{-2\alpha_0(r-\ell_{ij})}-2e^{-\alpha_0(r-\ell_{ij})}\big] ,\ \ \    r<r_c
\label{eq_U}
\end{equation}
where $\ell_{ij}$ is the average diameter of particles $i$ and $j$; the well-depth is $\varepsilon_0$ and the interaction range and cutoff parameters are $\alpha_0=33$ and $r_c=1.4 \ell_{ij}$. We use non-dimensionalized parameters throughout, 
the interaction strength is measured relative to $k_{\rm B}T$; the unit of time is $(m\bar\ell^2/k_{\rm B} T)^{1/2}=1$ where $\bar\ell$ is the mean particle diameter, and 
the volume fraction is $\phi=\pi N\overline{\ell^3}/(6V)$ where $N$ is the number of particles. 
We perform molecular dynamics with a Langevin thermostat and (non-dimensionalized) friction constant $\gamma_0=10$~\cite{razali2017effects}.
{(These Brownian simulations are computationally efficient and accurate enough to capture the essential physics of gelation, although they neglect hydrodynamic interactions, which do affect some aspects of gel structure~\cite{royall2015probing,graaf2018hydro}.)}
See Appendices
for further simulation details.

Starting from a random configuration at $\phi=0.2$, we prepare gels by simulating for a time $t_{\rm w}$, during which spinodal decomposition occurs.  
(We estimate the critical interaction strength for spinodal decomposition as $\varepsilon_0^*\approx3.13$~\cite{noro2000}, we take $\eps_0/\eps_0^*$ in the range $1.5$--$7$, always inside the spinodal.)
After this waiting time, we use the method of \cite{VezirovSM2015,CabrioluSM19} to impose a constant shear stress in the $xy$ plane  of non-dimensionalized strength $\sigma_0$ (measured relative to $k_{\rm B}T/\bar\ell^3$), allowing flow along the $x$-direction with Lees-Edwards boundary conditions.  The interaction strength is $\eprep$ during the preparation and $\varepsilon_0$ during the shear.  We mostly take $\eprep=\varepsilon_0$ which is the natural experimental condition, see however~\cite{Taylor2012}.  Unless otherwise stated we take $N=10^4$; all results are averaged over many independent runs (typically 150), to enable statistically robust conclusions.

\begin{figure}[t]
\centerline{\includegraphics[width=01\linewidth]{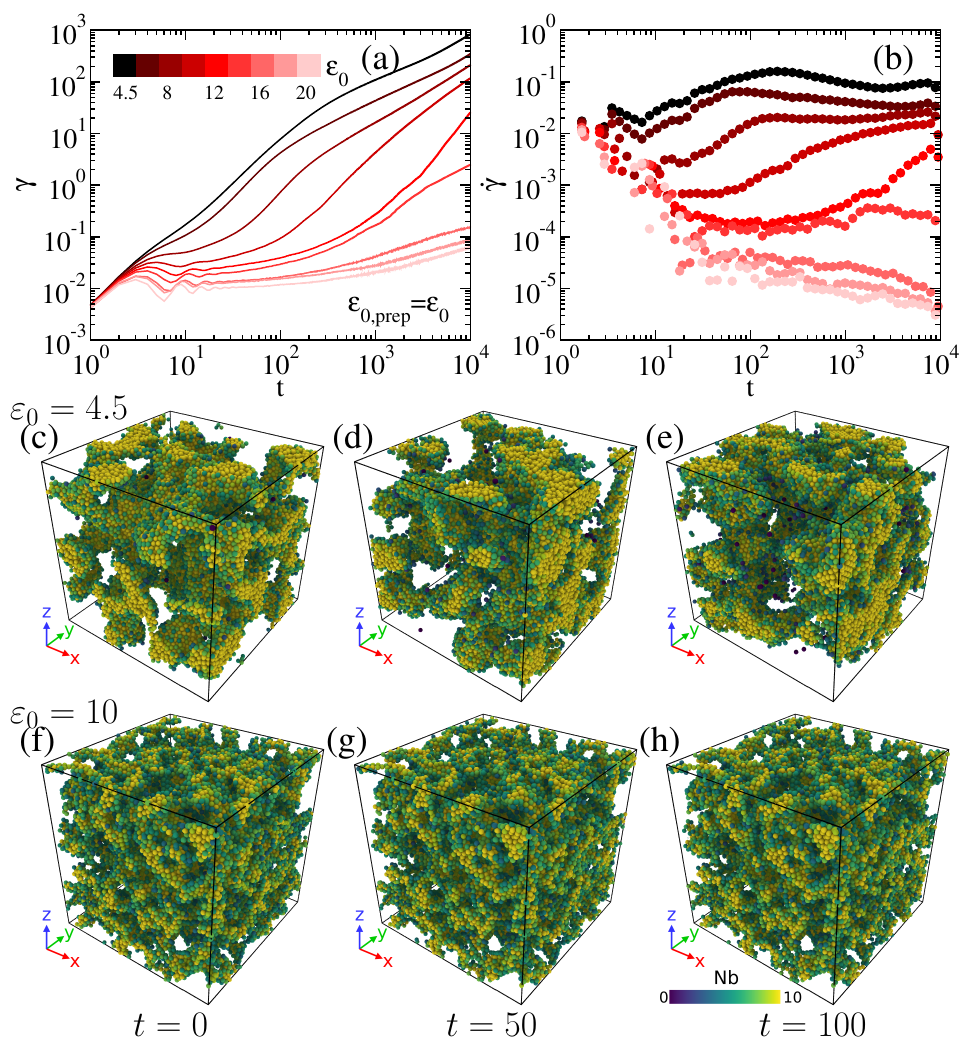}}
\caption{\label{fig1} (a) Strain against time for gels with different interaction strengths $\varepsilon_0$, at fixed $\sigma_0=1$. (b) Corresponding strain rate, estimated from finite differences of the strain.
(c,d,e)~Snapshots for a gel with $\varepsilon_0=4.5$ at different times, as shown.  Colours indicate the particle co-ordination $N_b$. (f,g,h)~Similar snapshots for $\varepsilon_0=10$.
}
\end{figure}

\emph{Flow, creep, and yielding}.
Fig.~\ref{fig1} shows the behavior of gels under constant stress $\sigma_0=1$, for various interaction strengths $\varepsilon_0$ (with $\eprep=\varepsilon_0$).
The average strain is plotted in Fig.~\ref{fig1}(a) as a function of the time $t$ since the start of shearing: for weak interactions, $\gamma$ increases smoothly from zero, because the imposed stress is large enough to break the arms of the gel.  For stronger interactions, there is an initial elastic deformation, followed by a plateau in $\gamma$, corresponding to a mechanically stable structure.  At longer times, the gels creep, leaving the plateau with an upward slope.  (The oscillations at early times are due to the constant-stress simulation method they do not affect the creep and yielding, see~\SMref.)  

Fig.~\ref{fig1}(b) shows the average shear rate.  For large $\varepsilon_0$, this decreases with time as the system enters the plateau, but the creeping motion means that it remains positive for all times.  For intermediate $\varepsilon_0$, the rate initially decreases but later increases again, which corresponds to yielding of the gel, after some creep.  For small $\varepsilon_0$ there is no plateau in the stress and the shear rate is relatively large throughout.  These dynamics are illustrated in Fig.~\ref{fig1}(c-e) for a weakly-interacting gel  which flows significantly in this time;  Fig.~\ref{fig1}(f-h) show a strongly interacting gel undergoing creep, with very small structural changes.
Note that weaker interactions allow spinodal decomposition to proceed more quickly, leading to thicker strands and larger pores at $t=0$, which strongly influence the future evolution.

\begin{figure}[!t]
\centerline{
 \includegraphics[width=1\linewidth]{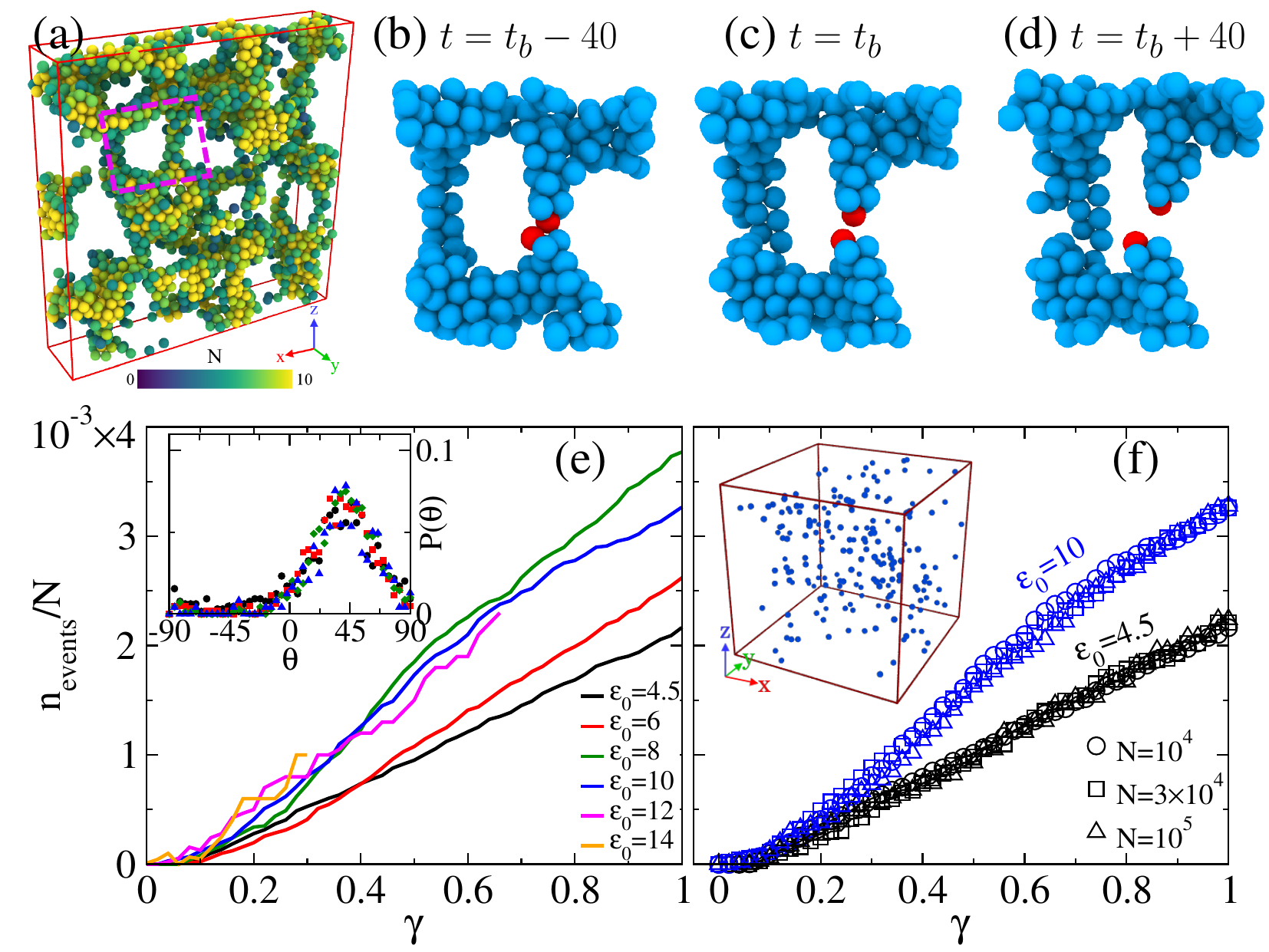}}
\caption{\label{fig2} (a)~Rendering of a slice through a gel, to visualize the strands. 
(b,c,d)~Expanded view of the boxed area highlighted in (a), showing a strand that breaks at time $t_{\rm b}$.
(e) Number of strand-breaking events $n_{\rm events}$ as a function of accumulated shear, normalized by the number of particles $N$.   Inset: distribution of orientations of the breaking strands.  (f) Number of strand-breaking events for two interaction strengths, varying the system size at fixed $\phi=0.2$.  Inset: locations in space of 300 representative events for a system with $N=10^5$.}
\end{figure}

\emph{Strand-breaking events}.
The gels are networks of strands, several particles thick.  These are broken as the gel yields~\cite{sprakel2011stress,lindstrom2012structures,van2018strand,verweij2019plasticity,KrisSMNeck23}.  
Figs. \ref{fig2}(a-d) show a section of the gel, focussing on one of the strands that breaks.
We have developed an algorithm for detection of such events, based on network topology, inspired by Ref.~\cite{tsurusawa2019direct}.  We identify a breaking strand as a pair of particles [colored red in Figs. \ref{fig2}(b-d)] for which the chemical distance (shortest bonded path) undergoes a sudden change, see \SM\ for details.

We analyse the statistics of thousands of such events, to understand yielding and flow.  Figs.~\ref{fig2}(e,f) plot the number of events, as a function of strain, showing of order $10^{-3}$ events per particle per unit strain, independent of system size. {This independence reflects that strand-breakage is a localised event; it stands in contrast to sheared athermal solids where plastic events near yielding are size-dependent~\cite{procaccia2017}.  We also find that}
 weaker gels have fewer events: since their arms are thicker, a single event has more impact on the structure; the arms may also be more flexible, allowing more strain without breakage.  
 
 The inset to Fig.~\ref{fig2}(e) shows the distribution of strand orientations as they break: we project the interparticle vector in the $xy$ plane, and define $\theta$ as the angle formed with the $x$-axis.  The distribution $P(\theta)$ is peaked around $45^\circ$, indicating that breaking strands are directed along the extensile direction of the shear flow.  The inset to Fig.~\ref{fig2}(f) shows that arm-breaking events are distributed homogeneously in space, with no sign of shear-banding or fracture.  Consistent with the very weak finite-size effects, this indicates a macroscopically ductile response, in contrast to fractal (non-reversible) gels~\cite{colombo2014stress,LeomachPRL14,SaintSM17ProteinGel}, whose behaviour is more brittle.

\begin{figure}[t]
\centerline{
\includegraphics[width=0.98\linewidth]{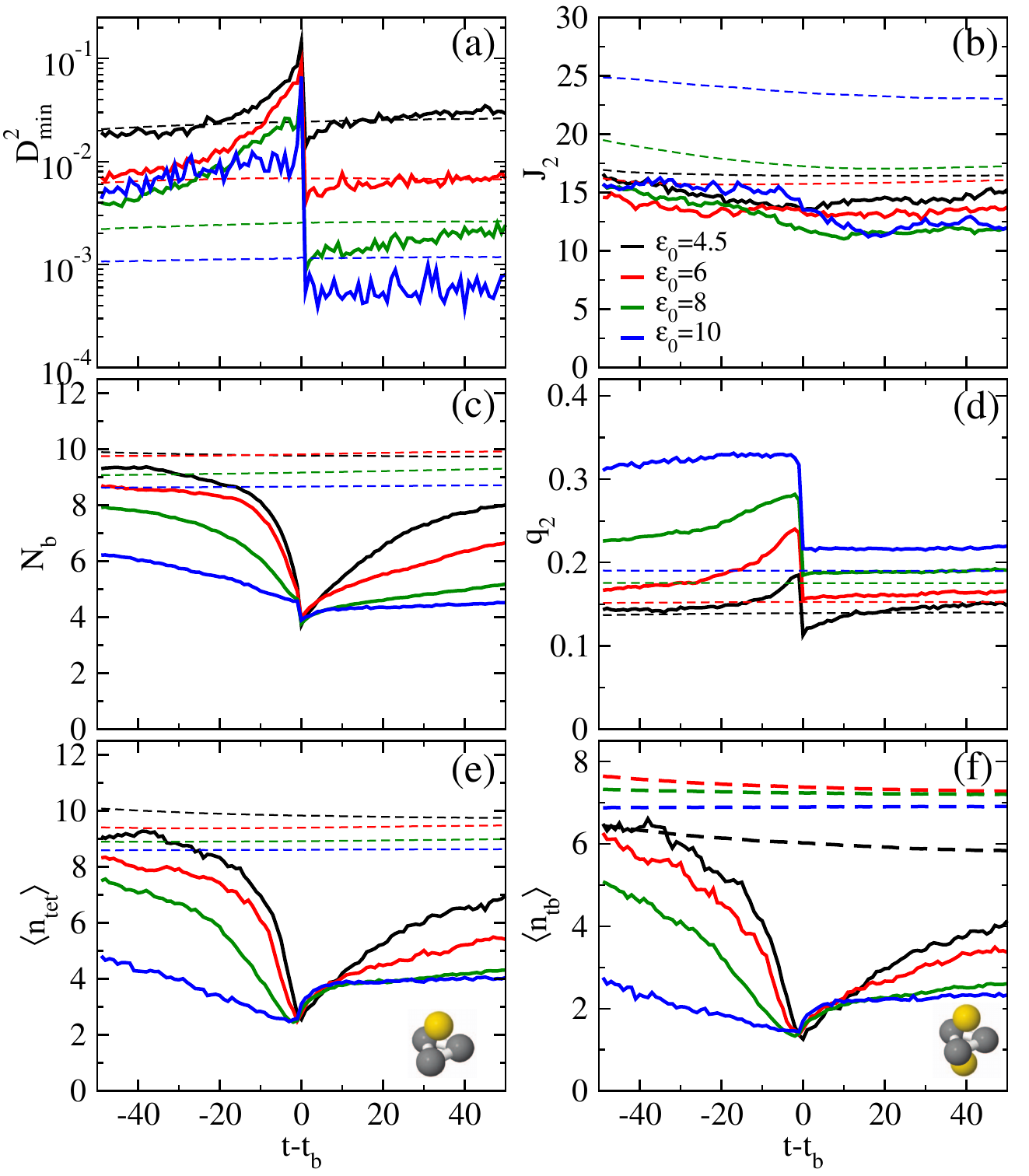}
}
\caption{\label{fig3} (a)~Averaged behaviour of non-affine displacement $D^2_{\rm min}$ for particles involved in strand-breaking events, as a function of the time relative to the event.  Dashed lines show the averaged behaviour for all particles.  Other panels show analogous results for: (b) stress anisotropy, (c) co-ordination number, (d) bond-order parameter $q_2$, (e,f) numbers of tetrahedra and triangular bipyramids in which the particles participate.
}
\end{figure}

\emph{Microscopic strand-breaking}.
We characterise local structure in strand-breaking events that occur before macroscopic yielding ($\gamma<1$ and $t<400$).  Detailed definitions of structural measurements are given in \SMref. Fig.~\ref{fig3}(a) shows the non-affine displacement $D^2_{\rm min}$~\cite{FalkPRE98D2min} over a short time period $\Delta t = 1$, for particles involved in strand breaking, such as those colored red in Fig.~\ref{fig2}.  As expected, $D^2_{\rm min}$ shows a peak when strands break.  Fig.~\ref{fig3} also shows the corresponding behaviour averaged over all particles, for comparison (dashed lines).

Note that $D^2_{\rm min}$ starts to increase significantly before the breakage event, especially for weaker interactions. We attribute this to a thinning and partial fluidisation of the arm before failure, as observed 
 for isolated single strands~\cite{KrisSMNeck23}.   Fig.~\ref{fig3}(a) shows the local Irving-Kirkwood~\cite{irving1950statistical} stress anisotropy $J_2$~\cite{KrisSMNeck23}: this  
 does not change rapidly at the breaking time but for larger $\eps_0$ one sees that particles have lower $J_2$ during strand breakage (solid lines), compared with the average (dashed).  This is again consistent with partial fluidisation relaxing the residual stresses in the arms, before breakage~\cite{KrisSMNeck23}.

We also characterise strands' local structure via their co-ordination numbers $N_b$, bond-orientational parameter $q_2$~\cite{tsurusawa2019direct}, and the topological cluster classification (TCC)~\cite{malins2013tcc} from which we extract the number of tetrahedra ($n_{\rm tet}$) and trigonal bipyramids ($n_{\rm tb}$)  in which particles participate.  We find a consistent trend -- strands that will break in the future already have non-typical structure for $t-t_b\approx -50$, characterised by fewer bonds ($N_b$), fewer locally-favoured structures ($n_{\rm tet},n_{\rm tb})$, and increasingly elongated bond orientations ($q_2$).  These pre-yielding effects are particularly pronounced for large $\varepsilon_0$, where the shear rates are also lowest.
After the breaking event ($t > t_b$), these quantities tend back towards their bulk values, presumably because the stretched arm is no longer in tension and can relax locally.  However, the TCC structures and $N_b$ both remain non-typical, presumably because particles which were involved in strand breakage are likely to remain near the surface of gel strands.

All together, the microscopic measurements of Fig.~\ref{fig2} indicate that strand-breaking is statistically \emph{predictable}, in that structural quantities before breakage differ from their typical values.  This offers a potential route towards material design and control of yielding.

\begin{figure}[!t]
\centerline{
\includegraphics[width=0.98\linewidth]{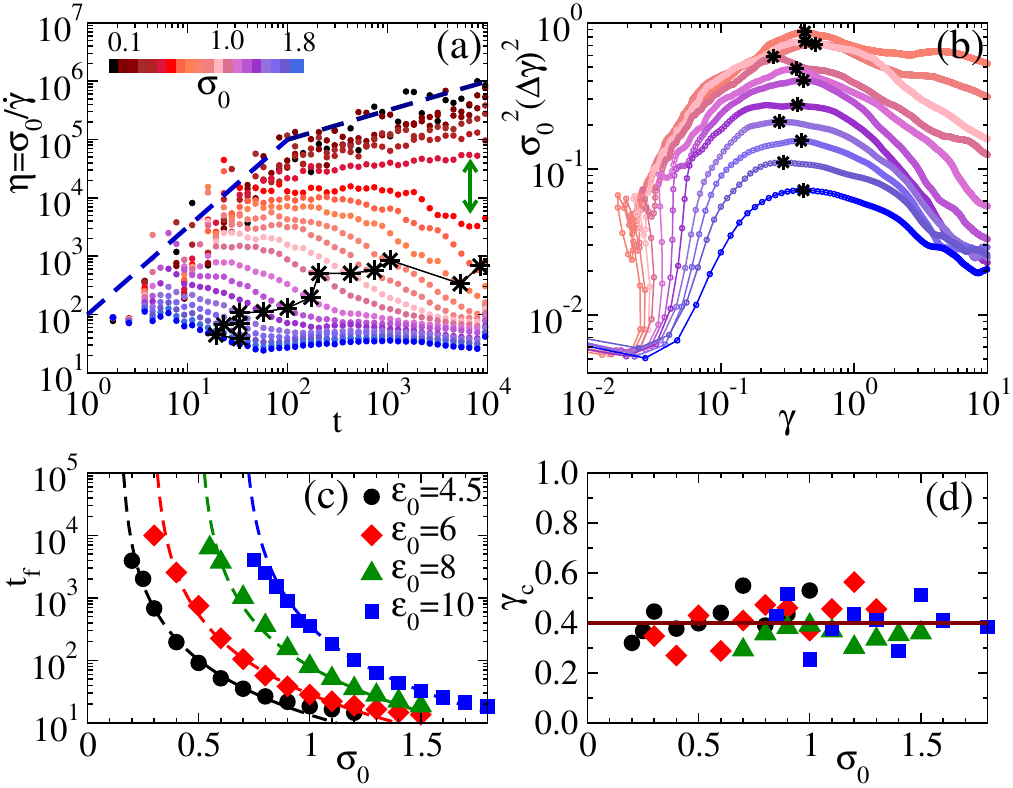}}
\caption{\label{fig4} (a)~Effective viscosity $\eta=\sigma_0/\dot{\gamma}$ against time, varying the imposed stress $\sigma_0$ at fixed $\varepsilon_0=10$.  Dashed lines indicate $\eta\sim t$ and $\eta \sim t^{0.5}$ for short and long times respectively.  
The green arrow indicates the bifurcation at $\sigma_0\approx0.7$.  (b)~Normalised strain fluctuations plotted against the average strain $\gamma$ for different $\sigma_0$ (color code shared with (a)). Black stars indicate the maxima which are interpreted as critical strains $\gamma_c$ (these points are also indicated in (a)).  (c)~Failure time $t_{\rm f}$ against imposed stress, with power law fits shown as dashed lines. (d)~Critical strains for all $\sigma_0,\varepsilon_0$ have $\gamma_c\approx 0.4$ (coloring shared with (c)).}
\end{figure}

\emph{Viscosity bifurcation}.
We now turn to macroscopic rheology, emphasizing that gels are far-from-equilibrium states so their structure tends to descend slowly in the energy landscape.  
For gels under shear, this can lead to a bifurcation similar to~\cite{BonnPRL2002,Coussot2002,BonnScience09}: aging strengthens the gel, which suppresses strand-breaking and allows further aging; on the other hand, if too many strands are broken then aging is interrupted and the gel weakens, promoting further breakage and yielding.
These effects are apparent in Fig. \ref{fig4}(a), which shows the effective viscosity, measured as the ratio of applied stress to measured shear rate, $\eta=\sigma_0/\dot{\gamma}$.  For small stress, the viscosity increases smoothly due to aging.  For larger stress, the viscosity initially increases, but strand-breaking events become important at later times, and it reduces.  This is the viscosity bifurcation.

For the time $t=10^4$ considered in Fig. \ref{fig4}(a), we identify a (weakly time-dependent) yield stress $\sigma_{y,t}=0.7$: an arrow separates samples above and below this value.
Following~\cite{CabrioluSM19}, we consider the sample-to-sample fluctuations of the strain as a function of $\gamma$: its normalised variance $(\Delta\gamma)^2$ shows a maximum at the yielding point, see Fig.~\ref{fig4}(b), and the Appendices.  The strain at this point is always $\gamma_c\approx 0.4$, and we define the failure time $t_{\rm f}$ as the average time taken to reach this strain.
We also varied $\varepsilon_0$: results for $t_{\rm f}$ are shown in Fig.\ \ref{fig4}(c).   The failure time increases as stress is reduced, it can be fitted as $t_{\rm f}\sim (\sigma_0-\sigma_{\infty})^{-\alpha}$ with  $\alpha\approx 2.4\pm0.1$ for all $\varepsilon_0$.  (The critical stress $\sigma_\infty$ increases with $\varepsilon_0$.) Fig.~\ref{fig4}(d) confirms that $\gamma_c$ depends weakly on $\eps_0$.

\begin{figure}[t]
\centerline{
\includegraphics[width=0.98\linewidth]{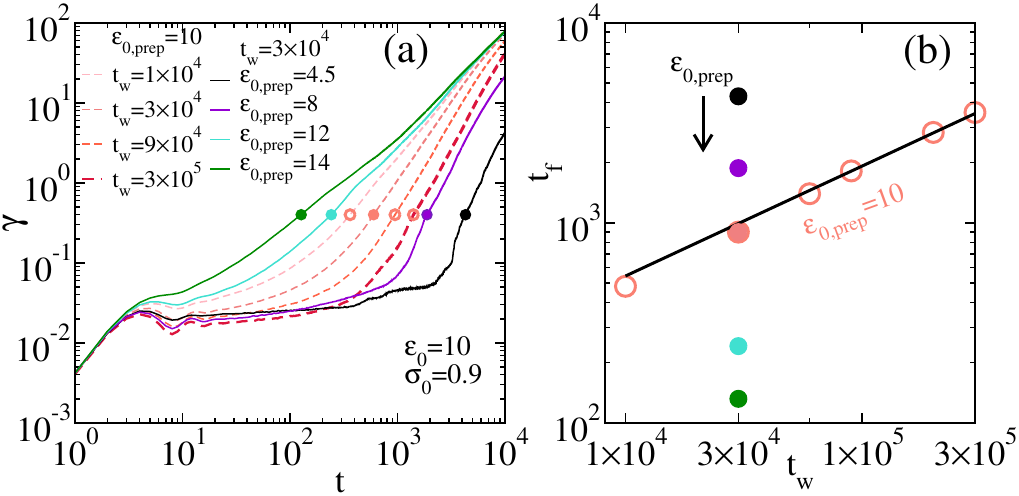}}
\caption{\label{fig5} (a) Strain against time for gels prepared with different $t_{\rm w}$ (all with $\eprep=10$) and for different $\eprep$ (all with $t_{\rm w}=3\times 10^4$).  Dots mark times $t_{\rm f}$ at which $\gamma=0.4$.
(b)~Failure times $t_{\rm f}$ increase with $t_{\rm w}$ and decrease with $\eprep$.  The straight line indicates $t_{\rm f}\sim t_{\rm w}^{1/2}$.  [Color coding for $\eprep$ is shared with (a).]}
\end{figure}

\emph{Effects of gel preparation}.  To further probe the competition between aging and shearing, we varied the waiting time $t_{\rm w}$ before the shear stress is applied.  This allows slightly thicker strands and larger voids to develop within the gel, see~\SMref.  We also varied $\eprep$ (interaction strength during the waiting time), while keeping fixed interaction strength $\eps$ during the shear.  (Recall that smaller $\eprep$ also leads to thicker gel strands.) Fig.~\ref{fig5} shows results for the strain, and for the failure time $t_{\rm f}$, extracted as in Fig.~\ref{fig4}.  The failure time increases significantly with $t_{\rm prep}$ because allowing coarsening to occur before shearing shifts the competition between aging and strand-breaking. This suppresses yielding and shifts the viscosity bifurcation to larger stress.  The data fits well to $t_{\rm f} \sim t_{\rm w}^{1/2}$ which differs from the (so-called) universal scaling $t_{\rm f} \sim t_{\rm w}$, found experimentally in fractal gels~\cite{WeitzPRL2000Universal}.
Reducing $\eprep$ has a similar effect to increasing $t_{\rm w}$: the resulting coarser gels are more resistant to shear.   These results are consistent with the expected phenomenology of the viscosity bifurcation~\cite{BonnPRL2002,Coussot2002,BonnScience09}. 

\textit{Conclusion} -- Our results provide a coherent picture of gel behavior on different length and time scales.  We linked local structure to strand-breaking events, and connected a ductile macroscopic response to a spatially homogeneous distribution of breaking events.   The aging of the local structure and the ductility are both linked to the relatively thick strands in these weak (depletion) gels, in contrast to fractal gels with thin strands~\cite{WeitzPRL2000Universal,colombo2014stress}.  A more detailed comparison between simulations of these two classes of gel would be valuable, to distentangle generic features  from model-specific results.

 Our results also offer several opportunities for future progress.  By uncovering relationships between strand-breaking events and microscopic structure, we offer a route towards ``bottom-up'' design of particulate gels, through their microscopic interactions.  
It would also be interesting to relate strands' behaviour near the viscosity bifurcation to the theoretical models that describe these phenomena at macroscopic scale~\cite{Coussot2002,Fielding2014}.
Based on of Fig.~\ref{fig2}, one may also propose an elastoplastic model~\cite{nicolas2018deformation} to capture the interdependence of different strand breakage events, whose microscopic properties would be encapsulated by model parameters.
Finally, the dependence of failure time $t_{\rm f}$ on $\eprep$ and $t_{\rm w}$ offers interesting opportunities for control of gel properties via their preparation conditions, including non-trivial protocols involving time-dependent interactions~\cite{Taylor2012,Royall2012quench}, and combining measurements with feedback protocols~\cite{KlostaJCP13,Tang2016}.

\begin{acknowledgments}
We thank Suzanne Fielding, Malcolm Faers, Kris Thijssen,  Abraham Mauleon-Amieva, and Rui Cheng for helpful discussions. This work was supported by the EPSRC through grants EP/T031247/1 (HB and RLJ) and EP/T031077/1 (CPR and TBL).
\end{acknowledgments}

\bibliography{Bibliography.bib} 

\begin{thebibliography}{68}%
\makeatletter
\providecommand \@ifxundefined [1]{%
 \@ifx{#1\undefined}
}%
\providecommand \@ifnum [1]{%
 \ifnum #1\expandafter \@firstoftwo
 \else \expandafter \@secondoftwo
 \fi
}%
\providecommand \@ifx [1]{%
 \ifx #1\expandafter \@firstoftwo
 \else \expandafter \@secondoftwo
 \fi
}%
\providecommand \natexlab [1]{#1}%
\providecommand \enquote  [1]{``#1''}%
\providecommand \bibnamefont  [1]{#1}%
\providecommand \bibfnamefont [1]{#1}%
\providecommand \citenamefont [1]{#1}%
\providecommand \href@noop [0]{\@secondoftwo}%
\providecommand \href [0]{\begingroup \@sanitize@url \@href}%
\providecommand \@href[1]{\@@startlink{#1}\@@href}%
\providecommand \@@href[1]{\endgroup#1\@@endlink}%
\providecommand \@sanitize@url [0]{\catcode `\\12\catcode `\$12\catcode
  `\&12\catcode `\#12\catcode `\^12\catcode `\_12\catcode `\%12\relax}%
\providecommand \@@startlink[1]{}%
\providecommand \@@endlink[0]{}%
\providecommand \url  [0]{\begingroup\@sanitize@url \@url }%
\providecommand \@url [1]{\endgroup\@href {#1}{\urlprefix }}%
\providecommand \urlprefix  [0]{URL }%
\providecommand \Eprint [0]{\href }%
\providecommand \doibase [0]{https://doi.org/}%
\providecommand \selectlanguage [0]{\@gobble}%
\providecommand \bibinfo  [0]{\@secondoftwo}%
\providecommand \bibfield  [0]{\@secondoftwo}%
\providecommand \translation [1]{[#1]}%
\providecommand \BibitemOpen [0]{}%
\providecommand \bibitemStop [0]{}%
\providecommand \bibitemNoStop [0]{.\EOS\space}%
\providecommand \EOS [0]{\spacefactor3000\relax}%
\providecommand \BibitemShut  [1]{\csname bibitem#1\endcsname}%
\let\auto@bib@innerbib\@empty
\bibitem [{\citenamefont {Zaccarelli}(2007)}]{zaccarelli2007colloidal}%
  \BibitemOpen
  \bibfield  {author} {\bibinfo {author} {\bibfnamefont {E.}~\bibnamefont
  {Zaccarelli}},\ }\bibfield  {title} {\bibinfo {title} {Colloidal gels:
  equilibrium and non-equilibrium routes},\ }\href@noop {} {\bibfield
  {journal} {\bibinfo  {journal} {Journal of Physics: Condensed Matter}\
  }\textbf {\bibinfo {volume} {19}},\ \bibinfo {pages} {323101} (\bibinfo
  {year} {2007})}\BibitemShut {NoStop}%
\bibitem [{\citenamefont {Royall}\ \emph {et~al.}(2021)\citenamefont {Royall},
  \citenamefont {Faers}, \citenamefont {Fussell},\ and\ \citenamefont
  {Hallett}}]{royall2021real}%
  \BibitemOpen
  \bibfield  {author} {\bibinfo {author} {\bibfnamefont {C.~P.}\ \bibnamefont
  {Royall}}, \bibinfo {author} {\bibfnamefont {M.~A.}\ \bibnamefont {Faers}},
  \bibinfo {author} {\bibfnamefont {S.~L.}\ \bibnamefont {Fussell}},\ and\
  \bibinfo {author} {\bibfnamefont {J.~E.}\ \bibnamefont {Hallett}},\
  }\bibfield  {title} {\bibinfo {title} {Real space analysis of colloidal gels:
  Triumphs, challenges and future directions},\ }\href@noop {} {\bibfield
  {journal} {\bibinfo  {journal} {Journal of Physics: Condensed Matter}\
  }\textbf {\bibinfo {volume} {33}},\ \bibinfo {pages} {453002} (\bibinfo
  {year} {2021})}\BibitemShut {NoStop}%
\bibitem [{\citenamefont {Cipelletti}\ and\ \citenamefont
  {Ramos}(2005)}]{cipelletti2005slow}%
  \BibitemOpen
  \bibfield  {author} {\bibinfo {author} {\bibfnamefont {L.}~\bibnamefont
  {Cipelletti}}\ and\ \bibinfo {author} {\bibfnamefont {L.}~\bibnamefont
  {Ramos}},\ }\bibfield  {title} {\bibinfo {title} {Slow dynamics in glassy
  soft matter},\ }\href@noop {} {\bibfield  {journal} {\bibinfo  {journal}
  {Journal of Physics: Condensed Matter}\ }\textbf {\bibinfo {volume} {17}},\
  \bibinfo {pages} {R253} (\bibinfo {year} {2005})}\BibitemShut {NoStop}%
\bibitem [{\citenamefont {Poon}(2002)}]{poon2002physics}%
  \BibitemOpen
  \bibfield  {author} {\bibinfo {author} {\bibfnamefont {W.}~\bibnamefont
  {Poon}},\ }\bibfield  {title} {\bibinfo {title} {The physics of a model
  colloid--polymer mixture},\ }\href@noop {} {\bibfield  {journal} {\bibinfo
  {journal} {Journal of Physics: Condensed Matter}\ }\textbf {\bibinfo {volume}
  {14}},\ \bibinfo {pages} {R859} (\bibinfo {year} {2002})}\BibitemShut
  {NoStop}%
\bibitem [{\citenamefont {Lu}\ \emph {et~al.}(2008)\citenamefont {Lu},
  \citenamefont {Zaccarelli}, \citenamefont {Ciulla}, \citenamefont
  {Schofield}, \citenamefont {Sciortino},\ and\ \citenamefont
  {Weitz}}]{lu2008gelation}%
  \BibitemOpen
  \bibfield  {author} {\bibinfo {author} {\bibfnamefont {P.~J.}\ \bibnamefont
  {Lu}}, \bibinfo {author} {\bibfnamefont {E.}~\bibnamefont {Zaccarelli}},
  \bibinfo {author} {\bibfnamefont {F.}~\bibnamefont {Ciulla}}, \bibinfo
  {author} {\bibfnamefont {A.~B.}\ \bibnamefont {Schofield}}, \bibinfo {author}
  {\bibfnamefont {F.}~\bibnamefont {Sciortino}},\ and\ \bibinfo {author}
  {\bibfnamefont {D.~A.}\ \bibnamefont {Weitz}},\ }\bibfield  {title} {\bibinfo
  {title} {Gelation of particles with short-range attraction},\ }\href@noop {}
  {\bibfield  {journal} {\bibinfo  {journal} {Nature}\ }\textbf {\bibinfo
  {volume} {453}},\ \bibinfo {pages} {499} (\bibinfo {year}
  {2008})}\BibitemShut {NoStop}%
\bibitem [{\citenamefont {Mezzenga}\ \emph {et~al.}(2005)\citenamefont
  {Mezzenga}, \citenamefont {Schurtenberger}, \citenamefont {Burbidge},\ and\
  \citenamefont {Michel}}]{mezzenga2005understanding}%
  \BibitemOpen
  \bibfield  {author} {\bibinfo {author} {\bibfnamefont {R.}~\bibnamefont
  {Mezzenga}}, \bibinfo {author} {\bibfnamefont {P.}~\bibnamefont
  {Schurtenberger}}, \bibinfo {author} {\bibfnamefont {A.}~\bibnamefont
  {Burbidge}},\ and\ \bibinfo {author} {\bibfnamefont {M.}~\bibnamefont
  {Michel}},\ }\bibfield  {title} {\bibinfo {title} {Understanding foods as
  soft materials},\ }\href@noop {} {\bibfield  {journal} {\bibinfo  {journal}
  {Nature materials}\ }\textbf {\bibinfo {volume} {4}},\ \bibinfo {pages} {729}
  (\bibinfo {year} {2005})}\BibitemShut {NoStop}%
\bibitem [{\citenamefont {Xiong}\ \emph {et~al.}(2019)\citenamefont {Xiong},
  \citenamefont {Yun}, \citenamefont {Qiu}, \citenamefont {Sun}, \citenamefont
  {Tang}, \citenamefont {He}, \citenamefont {Xiao}, \citenamefont {Chung},
  \citenamefont {Ng}, \citenamefont {Yan}, \citenamefont {Zhang}, \citenamefont
  {Wang},\ and\ \citenamefont {Li}}]{Xiong2019}%
  \BibitemOpen
  \bibfield  {author} {\bibinfo {author} {\bibfnamefont {Z.}~\bibnamefont
  {Xiong}}, \bibinfo {author} {\bibfnamefont {X.}~\bibnamefont {Yun}}, \bibinfo
  {author} {\bibfnamefont {L.}~\bibnamefont {Qiu}}, \bibinfo {author}
  {\bibfnamefont {Y.}~\bibnamefont {Sun}}, \bibinfo {author} {\bibfnamefont
  {B.}~\bibnamefont {Tang}}, \bibinfo {author} {\bibfnamefont {Z.}~\bibnamefont
  {He}}, \bibinfo {author} {\bibfnamefont {J.}~\bibnamefont {Xiao}}, \bibinfo
  {author} {\bibfnamefont {D.}~\bibnamefont {Chung}}, \bibinfo {author}
  {\bibfnamefont {T.~W.}\ \bibnamefont {Ng}}, \bibinfo {author} {\bibfnamefont
  {H.}~\bibnamefont {Yan}}, \bibinfo {author} {\bibfnamefont {R.}~\bibnamefont
  {Zhang}}, \bibinfo {author} {\bibfnamefont {X.}~\bibnamefont {Wang}},\ and\
  \bibinfo {author} {\bibfnamefont {D.}~\bibnamefont {Li}},\ }\bibfield
  {title} {\bibinfo {title} {A dynamic graphene oxide network enables spray
  printing of colloidal gels for high-performance micro-supercapacitors},\
  }\href@noop {} {\bibfield  {journal} {\bibinfo  {journal} {Advanced
  Materials}\ }\textbf {\bibinfo {volume} {31}},\ \bibinfo {pages} {1804434}
  (\bibinfo {year} {2019})}\BibitemShut {NoStop}%
\bibitem [{\citenamefont {Diba}\ \emph {et~al.}(2017)\citenamefont {Diba},
  \citenamefont {Wang}, \citenamefont {Kodger}, \citenamefont {Parsa},\ and\
  \citenamefont {Leeuwenburgh}}]{diba2017highly}%
  \BibitemOpen
  \bibfield  {author} {\bibinfo {author} {\bibfnamefont {M.}~\bibnamefont
  {Diba}}, \bibinfo {author} {\bibfnamefont {H.}~\bibnamefont {Wang}}, \bibinfo
  {author} {\bibfnamefont {T.~E.}\ \bibnamefont {Kodger}}, \bibinfo {author}
  {\bibfnamefont {S.}~\bibnamefont {Parsa}},\ and\ \bibinfo {author}
  {\bibfnamefont {S.~C.}\ \bibnamefont {Leeuwenburgh}},\ }\bibfield  {title}
  {\bibinfo {title} {Highly elastic and self-healing composite colloidal
  gels},\ }\href@noop {} {\bibfield  {journal} {\bibinfo  {journal} {Advanced
  materials}\ }\textbf {\bibinfo {volume} {29}},\ \bibinfo {pages} {1604672}
  (\bibinfo {year} {2017})}\BibitemShut {NoStop}%
\bibitem [{\citenamefont {Rose}\ \emph {et~al.}(2014)\citenamefont {Rose},
  \citenamefont {Prevoteau}, \citenamefont {Elzi\`{e}re}, \citenamefont
  {Hourdet}, \citenamefont {Marcellan},\ and\ \citenamefont
  {Leibler}}]{rose2014}%
  \BibitemOpen
  \bibfield  {author} {\bibinfo {author} {\bibfnamefont {S.}~\bibnamefont
  {Rose}}, \bibinfo {author} {\bibfnamefont {A.}~\bibnamefont {Prevoteau}},
  \bibinfo {author} {\bibfnamefont {P.}~\bibnamefont {Elzi\`{e}re}}, \bibinfo
  {author} {\bibfnamefont {D.}~\bibnamefont {Hourdet}}, \bibinfo {author}
  {\bibfnamefont {A.}~\bibnamefont {Marcellan}},\ and\ \bibinfo {author}
  {\bibfnamefont {L.}~\bibnamefont {Leibler}},\ }\bibfield  {title} {\bibinfo
  {title} {Nanoparticle solutions as adhesives for gels and biological
  tissues},\ }\href@noop {} {\bibfield  {journal} {\bibinfo  {journal}
  {Nature}\ }\textbf {\bibinfo {volume} {505}},\ \bibinfo {pages} {382}
  (\bibinfo {year} {2014})}\BibitemShut {NoStop}%
\bibitem [{\citenamefont {Bonn}\ \emph {et~al.}(2017)\citenamefont {Bonn},
  \citenamefont {Denn}, \citenamefont {Berthier}, \citenamefont {Divoux},\ and\
  \citenamefont {Manneville}}]{BonnRMP2017}%
  \BibitemOpen
  \bibfield  {author} {\bibinfo {author} {\bibfnamefont {D.}~\bibnamefont
  {Bonn}}, \bibinfo {author} {\bibfnamefont {M.~M.}\ \bibnamefont {Denn}},
  \bibinfo {author} {\bibfnamefont {L.}~\bibnamefont {Berthier}}, \bibinfo
  {author} {\bibfnamefont {T.}~\bibnamefont {Divoux}},\ and\ \bibinfo {author}
  {\bibfnamefont {S.}~\bibnamefont {Manneville}},\ }\bibfield  {title}
  {\bibinfo {title} {Yield stress materials in soft condensed matter},\ }\href
  {https://doi.org/10.1103/RevModPhys.89.035005} {\bibfield  {journal}
  {\bibinfo  {journal} {Rev. Mod. Phys.}\ }\textbf {\bibinfo {volume} {89}},\
  \bibinfo {pages} {035005} (\bibinfo {year} {2017})}\BibitemShut {NoStop}%
\bibitem [{\citenamefont {Divoux}\ \emph {et~al.}(2024)\citenamefont {Divoux},
  \citenamefont {Agoritsas}, \citenamefont {Aime}, \citenamefont {Barentin},
  \citenamefont {Barrat}, \citenamefont {Benzi}, \citenamefont {Berthier},
  \citenamefont {Bi}, \citenamefont {Biroli}, \citenamefont {Bonn},
  \citenamefont {Bourrianne}, \citenamefont {Bouzid}, \citenamefont {Del~Gado},
  \citenamefont {Delanoë-Ayari}, \citenamefont {Farain}, \citenamefont
  {Fielding}, \citenamefont {Fuchs}, \citenamefont {van~der Gucht},
  \citenamefont {Henkes}, \citenamefont {Jalaal}, \citenamefont {Joshi},
  \citenamefont {Lemaître}, \citenamefont {Leheny}, \citenamefont
  {Manneville}, \citenamefont {Martens}, \citenamefont {Poon}, \citenamefont
  {Popović}, \citenamefont {Procaccia}, \citenamefont {Ramos}, \citenamefont
  {Richards}, \citenamefont {Rogers}, \citenamefont {Rossi}, \citenamefont
  {Sbragaglia}, \citenamefont {Tarjus}, \citenamefont {Toschi}, \citenamefont
  {Trappe}, \citenamefont {Vermant}, \citenamefont {Wyart}, \citenamefont
  {Zamponi},\ and\ \citenamefont {Zare}}]{DivouxetalSMperspective24}%
  \BibitemOpen
  \bibfield  {author} {\bibinfo {author} {\bibfnamefont {T.}~\bibnamefont
  {Divoux}}, \bibinfo {author} {\bibfnamefont {E.}~\bibnamefont {Agoritsas}},
  \bibinfo {author} {\bibfnamefont {S.}~\bibnamefont {Aime}}, \bibinfo {author}
  {\bibfnamefont {C.}~\bibnamefont {Barentin}}, \bibinfo {author}
  {\bibfnamefont {J.-L.}\ \bibnamefont {Barrat}}, \bibinfo {author}
  {\bibfnamefont {R.}~\bibnamefont {Benzi}}, \bibinfo {author} {\bibfnamefont
  {L.}~\bibnamefont {Berthier}}, \bibinfo {author} {\bibfnamefont
  {D.}~\bibnamefont {Bi}}, \bibinfo {author} {\bibfnamefont {G.}~\bibnamefont
  {Biroli}}, \bibinfo {author} {\bibfnamefont {D.}~\bibnamefont {Bonn}},
  \bibinfo {author} {\bibfnamefont {P.}~\bibnamefont {Bourrianne}}, \bibinfo
  {author} {\bibfnamefont {M.}~\bibnamefont {Bouzid}}, \bibinfo {author}
  {\bibfnamefont {E.}~\bibnamefont {Del~Gado}}, \bibinfo {author}
  {\bibfnamefont {H.}~\bibnamefont {Delanoë-Ayari}}, \bibinfo {author}
  {\bibfnamefont {K.}~\bibnamefont {Farain}}, \bibinfo {author} {\bibfnamefont
  {S.}~\bibnamefont {Fielding}}, \bibinfo {author} {\bibfnamefont
  {M.}~\bibnamefont {Fuchs}}, \bibinfo {author} {\bibfnamefont
  {J.}~\bibnamefont {van~der Gucht}}, \bibinfo {author} {\bibfnamefont
  {S.}~\bibnamefont {Henkes}}, \bibinfo {author} {\bibfnamefont
  {M.}~\bibnamefont {Jalaal}}, \bibinfo {author} {\bibfnamefont {Y.~M.}\
  \bibnamefont {Joshi}}, \bibinfo {author} {\bibfnamefont {A.}~\bibnamefont
  {Lemaître}}, \bibinfo {author} {\bibfnamefont {R.~L.}\ \bibnamefont
  {Leheny}}, \bibinfo {author} {\bibfnamefont {S.}~\bibnamefont {Manneville}},
  \bibinfo {author} {\bibfnamefont {K.}~\bibnamefont {Martens}}, \bibinfo
  {author} {\bibfnamefont {W.~C.~K.}\ \bibnamefont {Poon}}, \bibinfo {author}
  {\bibfnamefont {M.}~\bibnamefont {Popović}}, \bibinfo {author}
  {\bibfnamefont {I.}~\bibnamefont {Procaccia}}, \bibinfo {author}
  {\bibfnamefont {L.}~\bibnamefont {Ramos}}, \bibinfo {author} {\bibfnamefont
  {J.~A.}\ \bibnamefont {Richards}}, \bibinfo {author} {\bibfnamefont
  {S.}~\bibnamefont {Rogers}}, \bibinfo {author} {\bibfnamefont
  {S.}~\bibnamefont {Rossi}}, \bibinfo {author} {\bibfnamefont
  {M.}~\bibnamefont {Sbragaglia}}, \bibinfo {author} {\bibfnamefont
  {G.}~\bibnamefont {Tarjus}}, \bibinfo {author} {\bibfnamefont
  {F.}~\bibnamefont {Toschi}}, \bibinfo {author} {\bibfnamefont
  {V.}~\bibnamefont {Trappe}}, \bibinfo {author} {\bibfnamefont
  {J.}~\bibnamefont {Vermant}}, \bibinfo {author} {\bibfnamefont
  {M.}~\bibnamefont {Wyart}}, \bibinfo {author} {\bibfnamefont
  {F.}~\bibnamefont {Zamponi}},\ and\ \bibinfo {author} {\bibfnamefont
  {D.}~\bibnamefont {Zare}},\ }\bibfield  {title} {\bibinfo {title}
  {Ductile-to-brittle transition and yielding in soft amorphous materials:
  perspectives and open questions},\ }\href
  {https://doi.org/10.1039/D3SM01740K} {\bibfield  {journal} {\bibinfo
  {journal} {Soft Matter}\ ,\ } (\bibinfo {year} {2024})}\BibitemShut {NoStop}%
\bibitem [{\citenamefont {Berthier}\ \emph {et~al.}(2024)\citenamefont
  {Berthier}, \citenamefont {Biroli}, \citenamefont {Manning},\ and\
  \citenamefont {Zamponi}}]{berthier2024yielding}%
  \BibitemOpen
  \bibfield  {author} {\bibinfo {author} {\bibfnamefont {L.}~\bibnamefont
  {Berthier}}, \bibinfo {author} {\bibfnamefont {G.}~\bibnamefont {Biroli}},
  \bibinfo {author} {\bibfnamefont {M.~L.}\ \bibnamefont {Manning}},\ and\
  \bibinfo {author} {\bibfnamefont {F.}~\bibnamefont {Zamponi}},\ }\href
  {https://arxiv.org/abs/2401.09385} {\bibinfo {title} {Yielding and plasticity
  in amorphous solids}} (\bibinfo {year} {2024}),\ \Eprint
  {https://arxiv.org/abs/2401.09385} {arXiv:2401.09385 [cond-mat.stat-mech]}
  \BibitemShut {NoStop}%
\bibitem [{\citenamefont {Cochran}\ \emph {et~al.}(2024)\citenamefont
  {Cochran}, \citenamefont {Callaghan}, \citenamefont {Caven},\ and\
  \citenamefont {Fielding}}]{cochran2024}%
  \BibitemOpen
  \bibfield  {author} {\bibinfo {author} {\bibfnamefont {J.~O.}\ \bibnamefont
  {Cochran}}, \bibinfo {author} {\bibfnamefont {G.~L.}\ \bibnamefont
  {Callaghan}}, \bibinfo {author} {\bibfnamefont {M.~J.~G.}\ \bibnamefont
  {Caven}},\ and\ \bibinfo {author} {\bibfnamefont {S.~M.}\ \bibnamefont
  {Fielding}},\ }\bibfield  {title} {\bibinfo {title} {Slow fatigue and highly
  delayed yielding via shear banding in oscillatory shear},\ }\href@noop {}
  {\bibfield  {journal} {\bibinfo  {journal} {Phys. Rev. Lett.}\ }\textbf
  {\bibinfo {volume} {132}},\ \bibinfo {pages} {168202} (\bibinfo {year}
  {2024})}\BibitemShut {NoStop}%
\bibitem [{\citenamefont {Bonn}\ and\ \citenamefont
  {Denn}(2009)}]{BonnScience09}%
  \BibitemOpen
  \bibfield  {author} {\bibinfo {author} {\bibfnamefont {D.}~\bibnamefont
  {Bonn}}\ and\ \bibinfo {author} {\bibfnamefont {M.~M.}\ \bibnamefont
  {Denn}},\ }\bibfield  {title} {\bibinfo {title} {Yield stress fluids slowly
  yield to analysis},\ }\href {https://doi.org/10.1126/science.1174217}
  {\bibfield  {journal} {\bibinfo  {journal} {Science}\ }\textbf {\bibinfo
  {volume} {324}},\ \bibinfo {pages} {1401} (\bibinfo {year}
  {2009})}\BibitemShut {NoStop}%
\bibitem [{\citenamefont {Bouzid}\ \emph {et~al.}(2017)\citenamefont {Bouzid},
  \citenamefont {Colombo}, \citenamefont {Barbosa},\ and\ \citenamefont
  {Del~Gado}}]{bouzid2017elastically}%
  \BibitemOpen
  \bibfield  {author} {\bibinfo {author} {\bibfnamefont {M.}~\bibnamefont
  {Bouzid}}, \bibinfo {author} {\bibfnamefont {J.}~\bibnamefont {Colombo}},
  \bibinfo {author} {\bibfnamefont {L.~V.}\ \bibnamefont {Barbosa}},\ and\
  \bibinfo {author} {\bibfnamefont {E.}~\bibnamefont {Del~Gado}},\ }\bibfield
  {title} {\bibinfo {title} {Elastically driven intermittent microscopic
  dynamics in soft solids},\ }\href@noop {} {\bibfield  {journal} {\bibinfo
  {journal} {Nature communications}\ }\textbf {\bibinfo {volume} {8}},\
  \bibinfo {pages} {1} (\bibinfo {year} {2017})}\BibitemShut {NoStop}%
\bibitem [{\citenamefont {Aime}\ \emph {et~al.}(2018)\citenamefont {Aime},
  \citenamefont {Ramos},\ and\ \citenamefont
  {Cipelletti}}]{aime2018microscopic}%
  \BibitemOpen
  \bibfield  {author} {\bibinfo {author} {\bibfnamefont {S.}~\bibnamefont
  {Aime}}, \bibinfo {author} {\bibfnamefont {L.}~\bibnamefont {Ramos}},\ and\
  \bibinfo {author} {\bibfnamefont {L.}~\bibnamefont {Cipelletti}},\ }\bibfield
   {title} {\bibinfo {title} {Microscopic dynamics and failure precursors of a
  gel under mechanical load},\ }\href@noop {} {\bibfield  {journal} {\bibinfo
  {journal} {Proceedings of the National Academy of Sciences}\ }\textbf
  {\bibinfo {volume} {115}},\ \bibinfo {pages} {3587} (\bibinfo {year}
  {2018})}\BibitemShut {NoStop}%
\bibitem [{\citenamefont {Lockwood}\ \emph {et~al.}(2024)\citenamefont
  {Lockwood}, \citenamefont {Agar},\ and\ \citenamefont
  {Fielding}}]{lockwood2024}%
  \BibitemOpen
  \bibfield  {author} {\bibinfo {author} {\bibfnamefont {H.~A.}\ \bibnamefont
  {Lockwood}}, \bibinfo {author} {\bibfnamefont {M.~H.}\ \bibnamefont {Agar}},\
  and\ \bibinfo {author} {\bibfnamefont {S.~M.}\ \bibnamefont {Fielding}},\
  }\bibfield  {title} {\bibinfo {title} {Power law creep and delayed failure of
  gels and fibrous materials under stress},\ }\href@noop {} {\bibfield
  {journal} {\bibinfo  {journal} {Soft Matter}\ }\textbf {\bibinfo {volume}
  {20}},\ \bibinfo {pages} {2474} (\bibinfo {year} {2024})}\BibitemShut
  {NoStop}%
\bibitem [{\citenamefont {Starrs}\ \emph {et~al.}(2002)\citenamefont {Starrs},
  \citenamefont {Poon}, \citenamefont {Hibberd},\ and\ \citenamefont
  {Robins}}]{Starrs2002}%
  \BibitemOpen
  \bibfield  {author} {\bibinfo {author} {\bibfnamefont {L.}~\bibnamefont
  {Starrs}}, \bibinfo {author} {\bibfnamefont {W.~C.~K.}\ \bibnamefont {Poon}},
  \bibinfo {author} {\bibfnamefont {D.~J.}\ \bibnamefont {Hibberd}},\ and\
  \bibinfo {author} {\bibfnamefont {M.~M.}\ \bibnamefont {Robins}},\ }\bibfield
   {title} {\bibinfo {title} {Collapse of transient gels in colloid-polymer
  mixtures},\ }\href@noop {} {\bibfield  {journal} {\bibinfo  {journal}
  {Journal of Physics: Condensed Matter}\ }\textbf {\bibinfo {volume} {14}},\
  \bibinfo {pages} {2485} (\bibinfo {year} {2002})}\BibitemShut {NoStop}%
\bibitem [{\citenamefont {Bartlett}\ \emph {et~al.}(2012)\citenamefont
  {Bartlett}, \citenamefont {Teece},\ and\ \citenamefont
  {Faers}}]{bartlett2012sudden}%
  \BibitemOpen
  \bibfield  {author} {\bibinfo {author} {\bibfnamefont {P.}~\bibnamefont
  {Bartlett}}, \bibinfo {author} {\bibfnamefont {L.~J.}\ \bibnamefont
  {Teece}},\ and\ \bibinfo {author} {\bibfnamefont {M.~A.}\ \bibnamefont
  {Faers}},\ }\bibfield  {title} {\bibinfo {title} {Sudden collapse of a
  colloidal gel},\ }\href@noop {} {\bibfield  {journal} {\bibinfo  {journal}
  {Physical Review E}\ }\textbf {\bibinfo {volume} {85}},\ \bibinfo {pages}
  {021404} (\bibinfo {year} {2012})}\BibitemShut {NoStop}%
\bibitem [{\citenamefont {Varga}\ \emph {et~al.}(2018)\citenamefont {Varga},
  \citenamefont {Hofmann},\ and\ \citenamefont {Swan}}]{Varga2018}%
  \BibitemOpen
  \bibfield  {author} {\bibinfo {author} {\bibfnamefont {Z.}~\bibnamefont
  {Varga}}, \bibinfo {author} {\bibfnamefont {J.~L.}\ \bibnamefont {Hofmann}},\
  and\ \bibinfo {author} {\bibfnamefont {J.~W.}\ \bibnamefont {Swan}},\
  }\bibfield  {title} {\bibinfo {title} {Modelling a hydrodynamic instability
  in freely settling colloidal gels},\ }\href
  {https://doi.org/10.1017/jfm.2018.725} {\bibfield  {journal} {\bibinfo
  {journal} {Journal of Fluid Mechanics}\ }\textbf {\bibinfo {volume} {856}},\
  \bibinfo {pages} {1014} (\bibinfo {year} {2018})}\BibitemShut {NoStop}%
\bibitem [{\citenamefont {Dinsmore}\ \emph {et~al.}(2006)\citenamefont
  {Dinsmore}, \citenamefont {Prasad}, \citenamefont {Wong},\ and\ \citenamefont
  {Weitz}}]{WeitzPRL06}%
  \BibitemOpen
  \bibfield  {author} {\bibinfo {author} {\bibfnamefont {A.~D.}\ \bibnamefont
  {Dinsmore}}, \bibinfo {author} {\bibfnamefont {V.}~\bibnamefont {Prasad}},
  \bibinfo {author} {\bibfnamefont {I.~Y.}\ \bibnamefont {Wong}},\ and\
  \bibinfo {author} {\bibfnamefont {D.~A.}\ \bibnamefont {Weitz}},\ }\bibfield
  {title} {\bibinfo {title} {Microscopic structure and elasticity of weakly
  aggregated colloidal gels},\ }\href
  {https://doi.org/10.1103/PhysRevLett.96.185502} {\bibfield  {journal}
  {\bibinfo  {journal} {Phys. Rev. Lett.}\ }\textbf {\bibinfo {volume} {96}},\
  \bibinfo {pages} {185502} (\bibinfo {year} {2006})}\BibitemShut {NoStop}%
\bibitem [{\citenamefont {Hsiao}\ \emph {et~al.}(2012)\citenamefont {Hsiao},
  \citenamefont {Newman}, \citenamefont {Glotzer},\ and\ \citenamefont
  {Solomon}}]{HsiaoPNAS12}%
  \BibitemOpen
  \bibfield  {author} {\bibinfo {author} {\bibfnamefont {L.~C.}\ \bibnamefont
  {Hsiao}}, \bibinfo {author} {\bibfnamefont {R.~S.}\ \bibnamefont {Newman}},
  \bibinfo {author} {\bibfnamefont {S.~C.}\ \bibnamefont {Glotzer}},\ and\
  \bibinfo {author} {\bibfnamefont {M.~J.}\ \bibnamefont {Solomon}},\
  }\bibfield  {title} {\bibinfo {title} {Role of isostaticity and load-bearing
  microstructure in the elasticity of yielded colloidal gels},\ }\href
  {https://doi.org/10.1073/pnas.1206742109} {\bibfield  {journal} {\bibinfo
  {journal} {Proceedings of the National Academy of Sciences}\ }\textbf
  {\bibinfo {volume} {109}},\ \bibinfo {pages} {16029} (\bibinfo {year}
  {2012})}\BibitemShut {NoStop}%
\bibitem [{\citenamefont {Tsurusawa}\ \emph {et~al.}(2019)\citenamefont
  {Tsurusawa}, \citenamefont {Leocmach}, \citenamefont {Russo},\ and\
  \citenamefont {Tanaka}}]{tsurusawa2019direct}%
  \BibitemOpen
  \bibfield  {author} {\bibinfo {author} {\bibfnamefont {H.}~\bibnamefont
  {Tsurusawa}}, \bibinfo {author} {\bibfnamefont {M.}~\bibnamefont {Leocmach}},
  \bibinfo {author} {\bibfnamefont {J.}~\bibnamefont {Russo}},\ and\ \bibinfo
  {author} {\bibfnamefont {H.}~\bibnamefont {Tanaka}},\ }\bibfield  {title}
  {\bibinfo {title} {Direct link between mechanical stability in gels and
  percolation of isostatic particles},\ }\href@noop {} {\bibfield  {journal}
  {\bibinfo  {journal} {Sci. Adv.}\ }\textbf {\bibinfo {volume} {5}},\ \bibinfo
  {pages} {eaav6090} (\bibinfo {year} {2019})}\BibitemShut {NoStop}%
\bibitem [{\citenamefont {Bantawa}\ \emph {et~al.}(2023)\citenamefont
  {Bantawa}, \citenamefont {Keshavarz}, \citenamefont {Geri}, \citenamefont
  {Bouzid}, \citenamefont {Divoux}, \citenamefont {McKinley},\ and\
  \citenamefont {Del~Gado}}]{bantawa2022hidden}%
  \BibitemOpen
  \bibfield  {author} {\bibinfo {author} {\bibfnamefont {M.}~\bibnamefont
  {Bantawa}}, \bibinfo {author} {\bibfnamefont {B.}~\bibnamefont {Keshavarz}},
  \bibinfo {author} {\bibfnamefont {M.}~\bibnamefont {Geri}}, \bibinfo {author}
  {\bibfnamefont {M.}~\bibnamefont {Bouzid}}, \bibinfo {author} {\bibfnamefont
  {T.}~\bibnamefont {Divoux}}, \bibinfo {author} {\bibfnamefont {G.~H.}\
  \bibnamefont {McKinley}},\ and\ \bibinfo {author} {\bibfnamefont
  {E.}~\bibnamefont {Del~Gado}},\ }\bibfield  {title} {\bibinfo {title} {The
  hidden hierarchical nature of soft particulate gels},\ }\href
  {https://doi.org/10.1038/s41567-023-01988-7} {\bibfield  {journal} {\bibinfo
  {journal} {Nature Physics}\ }\textbf {\bibinfo {volume} {19}},\ \bibinfo
  {pages} {1178} (\bibinfo {year} {2023})}\BibitemShut {NoStop}%
\bibitem [{\citenamefont {Nabizadeh}\ \emph {et~al.}(2024)\citenamefont
  {Nabizadeh}, \citenamefont {Nasirian}, \citenamefont {Li}, \citenamefont
  {Saraswat}, \citenamefont {Waheibi}, \citenamefont {Hsiao}, \citenamefont
  {Bi}, \citenamefont {Ravandi},\ and\ \citenamefont {Jamali}}]{JamaliPNAS24}%
  \BibitemOpen
  \bibfield  {author} {\bibinfo {author} {\bibfnamefont {M.}~\bibnamefont
  {Nabizadeh}}, \bibinfo {author} {\bibfnamefont {F.}~\bibnamefont {Nasirian}},
  \bibinfo {author} {\bibfnamefont {X.}~\bibnamefont {Li}}, \bibinfo {author}
  {\bibfnamefont {Y.}~\bibnamefont {Saraswat}}, \bibinfo {author}
  {\bibfnamefont {R.}~\bibnamefont {Waheibi}}, \bibinfo {author} {\bibfnamefont
  {L.~C.}\ \bibnamefont {Hsiao}}, \bibinfo {author} {\bibfnamefont
  {D.}~\bibnamefont {Bi}}, \bibinfo {author} {\bibfnamefont {B.}~\bibnamefont
  {Ravandi}},\ and\ \bibinfo {author} {\bibfnamefont {S.}~\bibnamefont
  {Jamali}},\ }\bibfield  {title} {\bibinfo {title} {Network physics of
  attractive colloidal gels: Resilience, rigidity, and phase diagram},\ }\href
  {https://doi.org/10.1073/pnas.2316394121} {\bibfield  {journal} {\bibinfo
  {journal} {Proceedings of the National Academy of Sciences}\ }\textbf
  {\bibinfo {volume} {121}},\ \bibinfo {pages} {e2316394121} (\bibinfo {year}
  {2024})}\BibitemShut {NoStop}%
\bibitem [{\citenamefont {Mangal}\ \emph
  {et~al.}(2024{\natexlab{a}})\citenamefont {Mangal}, \citenamefont
  {Nabizadeh},\ and\ \citenamefont {Jamali}}]{JamaliPRE24}%
  \BibitemOpen
  \bibfield  {author} {\bibinfo {author} {\bibfnamefont {D.}~\bibnamefont
  {Mangal}}, \bibinfo {author} {\bibfnamefont {M.}~\bibnamefont {Nabizadeh}},\
  and\ \bibinfo {author} {\bibfnamefont {S.}~\bibnamefont {Jamali}},\
  }\bibfield  {title} {\bibinfo {title} {Predicting yielding in attractive
  colloidal gels},\ }\href {https://doi.org/10.1103/PhysRevE.109.014602}
  {\bibfield  {journal} {\bibinfo  {journal} {Phys. Rev. E}\ }\textbf {\bibinfo
  {volume} {109}},\ \bibinfo {pages} {014602} (\bibinfo {year}
  {2024}{\natexlab{a}})}\BibitemShut {NoStop}%
\bibitem [{\citenamefont {van Doorn}\ \emph {et~al.}(2018)\citenamefont {van
  Doorn}, \citenamefont {Verweij}, \citenamefont {Sprakel},\ and\ \citenamefont
  {van~der Gucht}}]{van2018strand}%
  \BibitemOpen
  \bibfield  {author} {\bibinfo {author} {\bibfnamefont {J.~M.}\ \bibnamefont
  {van Doorn}}, \bibinfo {author} {\bibfnamefont {J.~E.}\ \bibnamefont
  {Verweij}}, \bibinfo {author} {\bibfnamefont {J.}~\bibnamefont {Sprakel}},\
  and\ \bibinfo {author} {\bibfnamefont {J.}~\bibnamefont {van~der Gucht}},\
  }\bibfield  {title} {\bibinfo {title} {Strand plasticity governs fatigue in
  colloidal gels},\ }\href@noop {} {\bibfield  {journal} {\bibinfo  {journal}
  {Physical review letters}\ }\textbf {\bibinfo {volume} {120}},\ \bibinfo
  {pages} {208005} (\bibinfo {year} {2018})}\BibitemShut {NoStop}%
\bibitem [{\citenamefont {Verweij}\ \emph {et~al.}(2019)\citenamefont
  {Verweij}, \citenamefont {Leermakers}, \citenamefont {Sprakel},\ and\
  \citenamefont {Van Der~Gucht}}]{verweij2019plasticity}%
  \BibitemOpen
  \bibfield  {author} {\bibinfo {author} {\bibfnamefont {J.~E.}\ \bibnamefont
  {Verweij}}, \bibinfo {author} {\bibfnamefont {F.~A.}\ \bibnamefont
  {Leermakers}}, \bibinfo {author} {\bibfnamefont {J.}~\bibnamefont
  {Sprakel}},\ and\ \bibinfo {author} {\bibfnamefont {J.}~\bibnamefont {Van
  Der~Gucht}},\ }\bibfield  {title} {\bibinfo {title} {Plasticity in colloidal
  gel strands},\ }\href@noop {} {\bibfield  {journal} {\bibinfo  {journal}
  {Soft Matter}\ }\textbf {\bibinfo {volume} {15}},\ \bibinfo {pages} {6447}
  (\bibinfo {year} {2019})}\BibitemShut {NoStop}%
\bibitem [{\citenamefont {Thijssen}\ \emph {et~al.}(2023)\citenamefont
  {Thijssen}, \citenamefont {Liverpool}, \citenamefont {Royall},\ and\
  \citenamefont {Jack}}]{KrisSMNeck23}%
  \BibitemOpen
  \bibfield  {author} {\bibinfo {author} {\bibfnamefont {K.}~\bibnamefont
  {Thijssen}}, \bibinfo {author} {\bibfnamefont {T.~B.}\ \bibnamefont
  {Liverpool}}, \bibinfo {author} {\bibfnamefont {C.~P.}\ \bibnamefont
  {Royall}},\ and\ \bibinfo {author} {\bibfnamefont {R.~L.}\ \bibnamefont
  {Jack}},\ }\bibfield  {title} {\bibinfo {title} {Necking and failure of a
  particulate gel strand: signatures of yielding on different length scales},\
  }\href {https://doi.org/10.1039/D3SM00681F} {\bibfield  {journal} {\bibinfo
  {journal} {Soft Matter}\ }\textbf {\bibinfo {volume} {19}},\ \bibinfo {pages}
  {7412} (\bibinfo {year} {2023})}\BibitemShut {NoStop}%
\bibitem [{\citenamefont {Sprakel}\ \emph {et~al.}(2011)\citenamefont
  {Sprakel}, \citenamefont {Lindstr{\"o}m}, \citenamefont {Kodger},\ and\
  \citenamefont {Weitz}}]{sprakel2011stress}%
  \BibitemOpen
  \bibfield  {author} {\bibinfo {author} {\bibfnamefont {J.}~\bibnamefont
  {Sprakel}}, \bibinfo {author} {\bibfnamefont {S.~B.}\ \bibnamefont
  {Lindstr{\"o}m}}, \bibinfo {author} {\bibfnamefont {T.~E.}\ \bibnamefont
  {Kodger}},\ and\ \bibinfo {author} {\bibfnamefont {D.~A.}\ \bibnamefont
  {Weitz}},\ }\bibfield  {title} {\bibinfo {title} {Stress enhancement in the
  delayed yielding of colloidal gels},\ }\href@noop {} {\bibfield  {journal}
  {\bibinfo  {journal} {Physical review letters}\ }\textbf {\bibinfo {volume}
  {106}},\ \bibinfo {pages} {248303} (\bibinfo {year} {2011})}\BibitemShut
  {NoStop}%
\bibitem [{\citenamefont {Lindstr{\"o}m}\ \emph {et~al.}(2012)\citenamefont
  {Lindstr{\"o}m}, \citenamefont {Kodger}, \citenamefont {Sprakel},\ and\
  \citenamefont {Weitz}}]{lindstrom2012structures}%
  \BibitemOpen
  \bibfield  {author} {\bibinfo {author} {\bibfnamefont {S.~B.}\ \bibnamefont
  {Lindstr{\"o}m}}, \bibinfo {author} {\bibfnamefont {T.~E.}\ \bibnamefont
  {Kodger}}, \bibinfo {author} {\bibfnamefont {J.}~\bibnamefont {Sprakel}},\
  and\ \bibinfo {author} {\bibfnamefont {D.~A.}\ \bibnamefont {Weitz}},\
  }\bibfield  {title} {\bibinfo {title} {Structures, stresses, and fluctuations
  in the delayed failure of colloidal gels},\ }\href@noop {} {\bibfield
  {journal} {\bibinfo  {journal} {Soft Matter}\ }\textbf {\bibinfo {volume}
  {8}},\ \bibinfo {pages} {3657} (\bibinfo {year} {2012})}\BibitemShut
  {NoStop}%
\bibitem [{\citenamefont {Colombo}\ and\ \citenamefont
  {Del~Gado}(2014)}]{colombo2014stress}%
  \BibitemOpen
  \bibfield  {author} {\bibinfo {author} {\bibfnamefont {J.}~\bibnamefont
  {Colombo}}\ and\ \bibinfo {author} {\bibfnamefont {E.}~\bibnamefont
  {Del~Gado}},\ }\bibfield  {title} {\bibinfo {title} {Stress localization,
  stiffening, and yielding in a model colloidal gel},\ }\href@noop {}
  {\bibfield  {journal} {\bibinfo  {journal} {Journal of rheology}\ }\textbf
  {\bibinfo {volume} {58}},\ \bibinfo {pages} {1089} (\bibinfo {year}
  {2014})}\BibitemShut {NoStop}%
\bibitem [{\citenamefont {Landrum}\ \emph {et~al.}(2016)\citenamefont
  {Landrum}, \citenamefont {Russel},\ and\ \citenamefont
  {Zia}}]{landrum2016delayed}%
  \BibitemOpen
  \bibfield  {author} {\bibinfo {author} {\bibfnamefont {B.~J.}\ \bibnamefont
  {Landrum}}, \bibinfo {author} {\bibfnamefont {W.~B.}\ \bibnamefont
  {Russel}},\ and\ \bibinfo {author} {\bibfnamefont {R.~N.}\ \bibnamefont
  {Zia}},\ }\bibfield  {title} {\bibinfo {title} {Delayed yield in colloidal
  gels: Creep, flow, and re-entrant solid regimes},\ }\href@noop {} {\bibfield
  {journal} {\bibinfo  {journal} {Journal of Rheology}\ }\textbf {\bibinfo
  {volume} {60}},\ \bibinfo {pages} {783} (\bibinfo {year} {2016})}\BibitemShut
  {NoStop}%
\bibitem [{\citenamefont {Saint-Michel}\ \emph {et~al.}(2017)\citenamefont
  {Saint-Michel}, \citenamefont {Gibaud},\ and\ \citenamefont
  {Manneville}}]{SaintSM17ProteinGel}%
  \BibitemOpen
  \bibfield  {author} {\bibinfo {author} {\bibfnamefont {B.}~\bibnamefont
  {Saint-Michel}}, \bibinfo {author} {\bibfnamefont {T.}~\bibnamefont
  {Gibaud}},\ and\ \bibinfo {author} {\bibfnamefont {S.}~\bibnamefont
  {Manneville}},\ }\bibfield  {title} {\bibinfo {title} {Predicting and
  assessing rupture in protein gels under oscillatory shear},\ }\href
  {https://doi.org/10.1039/C7SM00064B} {\bibfield  {journal} {\bibinfo
  {journal} {Soft Matter}\ }\textbf {\bibinfo {volume} {13}},\ \bibinfo {pages}
  {2643} (\bibinfo {year} {2017})}\BibitemShut {NoStop}%
\bibitem [{\citenamefont {Patrick~Royall}\ \emph {et~al.}(2008)\citenamefont
  {Patrick~Royall}, \citenamefont {Williams}, \citenamefont {Ohtsuka},\ and\
  \citenamefont {Tanaka}}]{patrick2008direct}%
  \BibitemOpen
  \bibfield  {author} {\bibinfo {author} {\bibfnamefont {C.}~\bibnamefont
  {Patrick~Royall}}, \bibinfo {author} {\bibfnamefont {S.~R.}\ \bibnamefont
  {Williams}}, \bibinfo {author} {\bibfnamefont {T.}~\bibnamefont {Ohtsuka}},\
  and\ \bibinfo {author} {\bibfnamefont {H.}~\bibnamefont {Tanaka}},\
  }\bibfield  {title} {\bibinfo {title} {Direct observation of a local
  structural mechanism for dynamic arrest},\ }\href@noop {} {\bibfield
  {journal} {\bibinfo  {journal} {Nature materials}\ }\textbf {\bibinfo
  {volume} {7}},\ \bibinfo {pages} {556} (\bibinfo {year} {2008})}\BibitemShut
  {NoStop}%
\bibitem [{\citenamefont {Griffiths}\ \emph {et~al.}(2017)\citenamefont
  {Griffiths}, \citenamefont {Turci},\ and\ \citenamefont
  {Royall}}]{griffiths2017local}%
  \BibitemOpen
  \bibfield  {author} {\bibinfo {author} {\bibfnamefont {S.}~\bibnamefont
  {Griffiths}}, \bibinfo {author} {\bibfnamefont {F.}~\bibnamefont {Turci}},\
  and\ \bibinfo {author} {\bibfnamefont {C.~P.}\ \bibnamefont {Royall}},\
  }\bibfield  {title} {\bibinfo {title} {Local structure of percolating gels at
  very low volume fractions},\ }\href@noop {} {\bibfield  {journal} {\bibinfo
  {journal} {The Journal of chemical physics}\ }\textbf {\bibinfo {volume}
  {146}},\ \bibinfo {pages} {014905} (\bibinfo {year} {2017})}\BibitemShut
  {NoStop}%
\bibitem [{\citenamefont {M{\"u}ller}\ \emph {et~al.}(2023)\citenamefont
  {M{\"u}ller}, \citenamefont {Isa},\ and\ \citenamefont
  {Vermant}}]{MullerNatCom2023}%
  \BibitemOpen
  \bibfield  {author} {\bibinfo {author} {\bibfnamefont {F.~J.}\ \bibnamefont
  {M{\"u}ller}}, \bibinfo {author} {\bibfnamefont {L.}~\bibnamefont {Isa}},\
  and\ \bibinfo {author} {\bibfnamefont {J.}~\bibnamefont {Vermant}},\
  }\bibfield  {title} {\bibinfo {title} {Toughening colloidal gels using rough
  building blocks},\ }\href {https://doi.org/10.1038/s41467-023-41098-9}
  {\bibfield  {journal} {\bibinfo  {journal} {Nature Communications}\ }\textbf
  {\bibinfo {volume} {14}},\ \bibinfo {pages} {5309} (\bibinfo {year}
  {2023})}\BibitemShut {NoStop}%
\bibitem [{\citenamefont {Mangal}\ \emph
  {et~al.}(2024{\natexlab{b}})\citenamefont {Mangal}, \citenamefont {Vera},
  \citenamefont {Aime},\ and\ \citenamefont {Jamali}}]{Mangal2024}%
  \BibitemOpen
  \bibfield  {author} {\bibinfo {author} {\bibfnamefont {D.}~\bibnamefont
  {Mangal}}, \bibinfo {author} {\bibfnamefont {G.~S.}\ \bibnamefont {Vera}},
  \bibinfo {author} {\bibfnamefont {S.}~\bibnamefont {Aime}},\ and\ \bibinfo
  {author} {\bibfnamefont {S.}~\bibnamefont {Jamali}},\ }\bibfield  {title}
  {\bibinfo {title} {Small variations in particle-level interactions lead to
  large structural heterogeneities in colloidal gels},\ }\href@noop {}
  {\bibfield  {journal} {\bibinfo  {journal} {Soft Matter}\ }\textbf {\bibinfo
  {volume} {20}},\ \bibinfo {pages} {4692} (\bibinfo {year}
  {2024}{\natexlab{b}})}\BibitemShut {NoStop}%
\bibitem [{\citenamefont {Coussot}\ \emph
  {et~al.}(2002{\natexlab{a}})\citenamefont {Coussot}, \citenamefont {Nguyen},
  \citenamefont {Huynh},\ and\ \citenamefont {Bonn}}]{BonnPRL2002}%
  \BibitemOpen
  \bibfield  {author} {\bibinfo {author} {\bibfnamefont {P.}~\bibnamefont
  {Coussot}}, \bibinfo {author} {\bibfnamefont {Q.~D.}\ \bibnamefont {Nguyen}},
  \bibinfo {author} {\bibfnamefont {H.~T.}\ \bibnamefont {Huynh}},\ and\
  \bibinfo {author} {\bibfnamefont {D.}~\bibnamefont {Bonn}},\ }\bibfield
  {title} {\bibinfo {title} {Avalanche behavior in yield stress fluids},\
  }\href {https://doi.org/10.1103/PhysRevLett.88.175501} {\bibfield  {journal}
  {\bibinfo  {journal} {Phys. Rev. Lett.}\ }\textbf {\bibinfo {volume} {88}},\
  \bibinfo {pages} {175501} (\bibinfo {year} {2002}{\natexlab{a}})}\BibitemShut
  {NoStop}%
\bibitem [{\citenamefont {Coussot}\ \emph
  {et~al.}(2002{\natexlab{b}})\citenamefont {Coussot}, \citenamefont {Nguyen},
  \citenamefont {Huynh},\ and\ \citenamefont {Bonn}}]{Coussot2002}%
  \BibitemOpen
  \bibfield  {author} {\bibinfo {author} {\bibfnamefont {P.}~\bibnamefont
  {Coussot}}, \bibinfo {author} {\bibfnamefont {Q.~D.}\ \bibnamefont {Nguyen}},
  \bibinfo {author} {\bibfnamefont {H.~T.}\ \bibnamefont {Huynh}},\ and\
  \bibinfo {author} {\bibfnamefont {D.}~\bibnamefont {Bonn}},\ }\bibfield
  {title} {\bibinfo {title} {Viscosity bifurcation in thixotropic, yielding
  fluids},\ }\href@noop {} {\bibfield  {journal} {\bibinfo  {journal} {Journal
  of Rheology}\ }\textbf {\bibinfo {volume} {46}},\ \bibinfo {pages} {573}
  (\bibinfo {year} {2002}{\natexlab{b}})}\BibitemShut {NoStop}%
\bibitem [{\citenamefont {Zia}\ \emph {et~al.}(2014)\citenamefont {Zia},
  \citenamefont {Landrum},\ and\ \citenamefont {Russel}}]{zia2014micro}%
  \BibitemOpen
  \bibfield  {author} {\bibinfo {author} {\bibfnamefont {R.~N.}\ \bibnamefont
  {Zia}}, \bibinfo {author} {\bibfnamefont {B.~J.}\ \bibnamefont {Landrum}},\
  and\ \bibinfo {author} {\bibfnamefont {W.~B.}\ \bibnamefont {Russel}},\
  }\bibfield  {title} {\bibinfo {title} {A micro-mechanical study of coarsening
  and rheology of colloidal gels: Cage building, cage hopping, and
  smoluchowski's ratchet},\ }\href@noop {} {\bibfield  {journal} {\bibinfo
  {journal} {Journal of Rheology}\ }\textbf {\bibinfo {volume} {58}},\ \bibinfo
  {pages} {1121} (\bibinfo {year} {2014})}\BibitemShut {NoStop}%
\bibitem [{\citenamefont {Koumakis}\ \emph {et~al.}(2015)\citenamefont
  {Koumakis}, \citenamefont {Moghimi}, \citenamefont {Besseling}, \citenamefont
  {Poon}, \citenamefont {Brady},\ and\ \citenamefont
  {Petekidis}}]{KoumakisSM15}%
  \BibitemOpen
  \bibfield  {author} {\bibinfo {author} {\bibfnamefont {N.}~\bibnamefont
  {Koumakis}}, \bibinfo {author} {\bibfnamefont {E.}~\bibnamefont {Moghimi}},
  \bibinfo {author} {\bibfnamefont {R.}~\bibnamefont {Besseling}}, \bibinfo
  {author} {\bibfnamefont {W.~C.~K.}\ \bibnamefont {Poon}}, \bibinfo {author}
  {\bibfnamefont {J.~F.}\ \bibnamefont {Brady}},\ and\ \bibinfo {author}
  {\bibfnamefont {G.}~\bibnamefont {Petekidis}},\ }\bibfield  {title} {\bibinfo
  {title} {Tuning colloidal gels by shear},\ }\href
  {https://doi.org/10.1039/C5SM00411J} {\bibfield  {journal} {\bibinfo
  {journal} {Soft Matter}\ }\textbf {\bibinfo {volume} {11}},\ \bibinfo {pages}
  {4640} (\bibinfo {year} {2015})}\BibitemShut {NoStop}%
\bibitem [{\citenamefont {Cipelletti}\ \emph {et~al.}(2000)\citenamefont
  {Cipelletti}, \citenamefont {Manley}, \citenamefont {Ball},\ and\
  \citenamefont {Weitz}}]{WeitzPRL2000Universal}%
  \BibitemOpen
  \bibfield  {author} {\bibinfo {author} {\bibfnamefont {L.}~\bibnamefont
  {Cipelletti}}, \bibinfo {author} {\bibfnamefont {S.}~\bibnamefont {Manley}},
  \bibinfo {author} {\bibfnamefont {R.~C.}\ \bibnamefont {Ball}},\ and\
  \bibinfo {author} {\bibfnamefont {D.~A.}\ \bibnamefont {Weitz}},\ }\bibfield
  {title} {\bibinfo {title} {Universal aging features in the restructuring of
  fractal colloidal gels},\ }\href
  {https://doi.org/10.1103/PhysRevLett.84.2275} {\bibfield  {journal} {\bibinfo
   {journal} {Phys. Rev. Lett.}\ }\textbf {\bibinfo {volume} {84}},\ \bibinfo
  {pages} {2275} (\bibinfo {year} {2000})}\BibitemShut {NoStop}%
\bibitem [{\citenamefont {Fenton}\ \emph {et~al.}(2023)\citenamefont {Fenton},
  \citenamefont {Padmanabhan}, \citenamefont {Ryu}, \citenamefont {Nguyen},
  \citenamefont {Zia},\ and\ \citenamefont {Helgeson}}]{Fenton2023}%
  \BibitemOpen
  \bibfield  {author} {\bibinfo {author} {\bibfnamefont {S.~M.}\ \bibnamefont
  {Fenton}}, \bibinfo {author} {\bibfnamefont {P.}~\bibnamefont {Padmanabhan}},
  \bibinfo {author} {\bibfnamefont {B.~K.}\ \bibnamefont {Ryu}}, \bibinfo
  {author} {\bibfnamefont {T.~T.~D.}\ \bibnamefont {Nguyen}}, \bibinfo {author}
  {\bibfnamefont {R.~N.}\ \bibnamefont {Zia}},\ and\ \bibinfo {author}
  {\bibfnamefont {M.~E.}\ \bibnamefont {Helgeson}},\ }\bibfield  {title}
  {\bibinfo {title} {Minimal conditions for solidification and thermal
  processing of colloidal gels},\ }\href@noop {} {\bibfield  {journal}
  {\bibinfo  {journal} {Proceedings of the National Academy of Sciences}\
  }\textbf {\bibinfo {volume} {120}},\ \bibinfo {pages} {e2215922120} (\bibinfo
  {year} {2023})}\BibitemShut {NoStop}%
\bibitem [{\citenamefont {Leocmach}\ \emph {et~al.}(2014)\citenamefont
  {Leocmach}, \citenamefont {Perge}, \citenamefont {Divoux},\ and\
  \citenamefont {Manneville}}]{LeomachPRL14}%
  \BibitemOpen
  \bibfield  {author} {\bibinfo {author} {\bibfnamefont {M.}~\bibnamefont
  {Leocmach}}, \bibinfo {author} {\bibfnamefont {C.}~\bibnamefont {Perge}},
  \bibinfo {author} {\bibfnamefont {T.}~\bibnamefont {Divoux}},\ and\ \bibinfo
  {author} {\bibfnamefont {S.}~\bibnamefont {Manneville}},\ }\bibfield  {title}
  {\bibinfo {title} {Creep and fracture of a protein gel under stress},\ }\href
  {https://doi.org/10.1103/PhysRevLett.113.038303} {\bibfield  {journal}
  {\bibinfo  {journal} {Phys. Rev. Lett.}\ }\textbf {\bibinfo {volume} {113}},\
  \bibinfo {pages} {038303} (\bibinfo {year} {2014})}\BibitemShut {NoStop}%
\bibitem [{\citenamefont {Taffs}\ \emph {et~al.}(2010)\citenamefont {Taffs},
  \citenamefont {Malins}, \citenamefont {Williams},\ and\ \citenamefont
  {Royall}}]{taffs2010}%
  \BibitemOpen
  \bibfield  {author} {\bibinfo {author} {\bibfnamefont {J.}~\bibnamefont
  {Taffs}}, \bibinfo {author} {\bibfnamefont {A.}~\bibnamefont {Malins}},
  \bibinfo {author} {\bibfnamefont {S.~R.}\ \bibnamefont {Williams}},\ and\
  \bibinfo {author} {\bibfnamefont {C.~P.}\ \bibnamefont {Royall}},\ }\bibfield
   {title} {\bibinfo {title} {A structural comparison of models of
  colloid-polymer mixtures},\ }\href@noop {} {\bibfield  {journal} {\bibinfo
  {journal} {J. Phys.: Condens. Matter}\ }\textbf {\bibinfo {volume} {22}},\
  \bibinfo {pages} {104119} (\bibinfo {year} {2010})}\BibitemShut {NoStop}%
\bibitem [{\citenamefont {Royall}\ \emph {et~al.}(2018)\citenamefont {Royall},
  \citenamefont {Williams},\ and\ \citenamefont
  {Tanaka}}]{royall2018vitrification}%
  \BibitemOpen
  \bibfield  {author} {\bibinfo {author} {\bibfnamefont {C.~P.}\ \bibnamefont
  {Royall}}, \bibinfo {author} {\bibfnamefont {S.~R.}\ \bibnamefont
  {Williams}},\ and\ \bibinfo {author} {\bibfnamefont {H.}~\bibnamefont
  {Tanaka}},\ }\bibfield  {title} {\bibinfo {title} {Vitrification and gelation
  in sticky spheres},\ }\href@noop {} {\bibfield  {journal} {\bibinfo
  {journal} {The Journal of chemical physics}\ }\textbf {\bibinfo {volume}
  {148}},\ \bibinfo {pages} {044501} (\bibinfo {year} {2018})}\BibitemShut
  {NoStop}%
\bibitem [{\citenamefont {Razali}\ \emph {et~al.}(2017)\citenamefont {Razali},
  \citenamefont {Fullerton}, \citenamefont {Turci}, \citenamefont {Hallett},
  \citenamefont {Jack},\ and\ \citenamefont {Royall}}]{razali2017effects}%
  \BibitemOpen
  \bibfield  {author} {\bibinfo {author} {\bibfnamefont {A.}~\bibnamefont
  {Razali}}, \bibinfo {author} {\bibfnamefont {C.~J.}\ \bibnamefont
  {Fullerton}}, \bibinfo {author} {\bibfnamefont {F.}~\bibnamefont {Turci}},
  \bibinfo {author} {\bibfnamefont {J.~E.}\ \bibnamefont {Hallett}}, \bibinfo
  {author} {\bibfnamefont {R.~L.}\ \bibnamefont {Jack}},\ and\ \bibinfo
  {author} {\bibfnamefont {C.~P.}\ \bibnamefont {Royall}},\ }\bibfield  {title}
  {\bibinfo {title} {Effects of vertical confinement on gelation and
  sedimentation of colloids},\ }\href@noop {} {\bibfield  {journal} {\bibinfo
  {journal} {Soft Matter}\ }\textbf {\bibinfo {volume} {13}},\ \bibinfo {pages}
  {3230} (\bibinfo {year} {2017})}\BibitemShut {NoStop}%
\bibitem [{\citenamefont {Royall}\ \emph {et~al.}(2015)\citenamefont {Royall},
  \citenamefont {Eggers}, \citenamefont {Furukawa},\ and\ \citenamefont
  {Tanaka}}]{royall2015probing}%
  \BibitemOpen
  \bibfield  {author} {\bibinfo {author} {\bibfnamefont {C.~P.}\ \bibnamefont
  {Royall}}, \bibinfo {author} {\bibfnamefont {J.}~\bibnamefont {Eggers}},
  \bibinfo {author} {\bibfnamefont {A.}~\bibnamefont {Furukawa}},\ and\
  \bibinfo {author} {\bibfnamefont {H.}~\bibnamefont {Tanaka}},\ }\bibfield
  {title} {\bibinfo {title} {Probing colloidal gels at multiple length scales:
  The role of hydrodynamics},\ }\href@noop {} {\bibfield  {journal} {\bibinfo
  {journal} {Phys. Rev. Lett.}\ }\textbf {\bibinfo {volume} {114}},\ \bibinfo
  {pages} {258302} (\bibinfo {year} {2015})}\BibitemShut {NoStop}%
\bibitem [{\citenamefont {de~Graaf}\ \emph {et~al.}(2019)\citenamefont
  {de~Graaf}, \citenamefont {Poon}, \citenamefont {Haughey},\ and\
  \citenamefont {Hermes}}]{graaf2018hydro}%
  \BibitemOpen
  \bibfield  {author} {\bibinfo {author} {\bibfnamefont {J.}~\bibnamefont
  {de~Graaf}}, \bibinfo {author} {\bibfnamefont {W.~C.~K.}\ \bibnamefont
  {Poon}}, \bibinfo {author} {\bibfnamefont {M.~J.}\ \bibnamefont {Haughey}},\
  and\ \bibinfo {author} {\bibfnamefont {M.}~\bibnamefont {Hermes}},\
  }\bibfield  {title} {\bibinfo {title} {Hydrodynamics strongly affect the
  dynamics of colloidal gelation but not gel structure},\ }\href@noop {}
  {\bibfield  {journal} {\bibinfo  {journal} {Soft Matter}\ }\textbf {\bibinfo
  {volume} {15}},\ \bibinfo {pages} {10} (\bibinfo {year} {2019})}\BibitemShut
  {NoStop}%
\bibitem [{\citenamefont {Noro}\ and\ \citenamefont
  {Frenkel}(2000)}]{noro2000}%
  \BibitemOpen
  \bibfield  {author} {\bibinfo {author} {\bibfnamefont {M.~G.}\ \bibnamefont
  {Noro}}\ and\ \bibinfo {author} {\bibfnamefont {D.}~\bibnamefont {Frenkel}},\
  }\bibfield  {title} {\bibinfo {title} {Extended corresponding-states behavior
  for particles with variable range attractions},\ }\href@noop {} {\bibfield
  {journal} {\bibinfo  {journal} {J. Chem. Phys.}\ }\textbf {\bibinfo {volume}
  {113}},\ \bibinfo {pages} {2941} (\bibinfo {year} {2000})}\BibitemShut
  {NoStop}%
\bibitem [{\citenamefont {Vezirov}\ \emph {et~al.}(2015)\citenamefont
  {Vezirov}, \citenamefont {Gerloff},\ and\ \citenamefont
  {Klapp}}]{VezirovSM2015}%
  \BibitemOpen
  \bibfield  {author} {\bibinfo {author} {\bibfnamefont {T.~A.}\ \bibnamefont
  {Vezirov}}, \bibinfo {author} {\bibfnamefont {S.}~\bibnamefont {Gerloff}},\
  and\ \bibinfo {author} {\bibfnamefont {S.~H.~L.}\ \bibnamefont {Klapp}},\
  }\bibfield  {title} {\bibinfo {title} {Manipulating shear-induced
  non-equilibrium transitions in colloidal films by feedback control},\ }\href
  {https://doi.org/10.1039/C4SM01414F} {\bibfield  {journal} {\bibinfo
  {journal} {Soft Matter}\ }\textbf {\bibinfo {volume} {11}},\ \bibinfo {pages}
  {406} (\bibinfo {year} {2015})}\BibitemShut {NoStop}%
\bibitem [{\citenamefont {Cabriolu}\ \emph {et~al.}(2019)\citenamefont
  {Cabriolu}, \citenamefont {Horbach}, \citenamefont {Chaudhuri},\ and\
  \citenamefont {Martens}}]{CabrioluSM19}%
  \BibitemOpen
  \bibfield  {author} {\bibinfo {author} {\bibfnamefont {R.}~\bibnamefont
  {Cabriolu}}, \bibinfo {author} {\bibfnamefont {J.}~\bibnamefont {Horbach}},
  \bibinfo {author} {\bibfnamefont {P.}~\bibnamefont {Chaudhuri}},\ and\
  \bibinfo {author} {\bibfnamefont {K.}~\bibnamefont {Martens}},\ }\bibfield
  {title} {\bibinfo {title} {Precursors of fluidisation in the creep response
  of a soft glass},\ }\href {https://doi.org/10.1039/C8SM01432A} {\bibfield
  {journal} {\bibinfo  {journal} {Soft Matter}\ }\textbf {\bibinfo {volume}
  {15}},\ \bibinfo {pages} {415} (\bibinfo {year} {2019})}\BibitemShut
  {NoStop}%
\bibitem [{\citenamefont {Taylor}\ \emph {et~al.}(2012)\citenamefont {Taylor},
  \citenamefont {Evans},\ and\ \citenamefont {Royall}}]{Taylor2012}%
  \BibitemOpen
  \bibfield  {author} {\bibinfo {author} {\bibfnamefont {S.~L.}\ \bibnamefont
  {Taylor}}, \bibinfo {author} {\bibfnamefont {R.}~\bibnamefont {Evans}},\ and\
  \bibinfo {author} {\bibfnamefont {C.~P.}\ \bibnamefont {Royall}},\ }\bibfield
   {title} {\bibinfo {title} {Temperature as an external field for
  colloid--polymer mixtures: quenching by heating and melting by cooling},\
  }\href@noop {} {\bibfield  {journal} {\bibinfo  {journal} {J. Phys.: Cond.
  Matt.}\ }\textbf {\bibinfo {volume} {24}},\ \bibinfo {pages} {464128}
  (\bibinfo {year} {2012})}\BibitemShut {NoStop}%
\bibitem [{\citenamefont {Procaccia}\ \emph {et~al.}(2017)\citenamefont
  {Procaccia}, \citenamefont {Rainone},\ and\ \citenamefont
  {Singh}}]{procaccia2017}%
  \BibitemOpen
  \bibfield  {author} {\bibinfo {author} {\bibfnamefont {I.}~\bibnamefont
  {Procaccia}}, \bibinfo {author} {\bibfnamefont {C.}~\bibnamefont {Rainone}},\
  and\ \bibinfo {author} {\bibfnamefont {M.}~\bibnamefont {Singh}},\ }\bibfield
   {title} {\bibinfo {title} {Mechanical failure in amorphous solids:
  Scale-free spinodal criticality},\ }\href@noop {} {\bibfield  {journal}
  {\bibinfo  {journal} {Phys. Rev. E}\ }\textbf {\bibinfo {volume} {96}},\
  \bibinfo {pages} {032907} (\bibinfo {year} {2017})}\BibitemShut {NoStop}%
\bibitem [{\citenamefont {Falk}\ and\ \citenamefont
  {Langer}(1998)}]{FalkPRE98D2min}%
  \BibitemOpen
  \bibfield  {author} {\bibinfo {author} {\bibfnamefont {M.~L.}\ \bibnamefont
  {Falk}}\ and\ \bibinfo {author} {\bibfnamefont {J.~S.}\ \bibnamefont
  {Langer}},\ }\bibfield  {title} {\bibinfo {title} {Dynamics of viscoplastic
  deformation in amorphous solids},\ }\href
  {https://doi.org/10.1103/PhysRevE.57.7192} {\bibfield  {journal} {\bibinfo
  {journal} {Phys. Rev. E}\ }\textbf {\bibinfo {volume} {57}},\ \bibinfo
  {pages} {7192} (\bibinfo {year} {1998})}\BibitemShut {NoStop}%
\bibitem [{\citenamefont {Irving}\ and\ \citenamefont
  {Kirkwood}(1950)}]{irving1950statistical}%
  \BibitemOpen
  \bibfield  {author} {\bibinfo {author} {\bibfnamefont {J.}~\bibnamefont
  {Irving}}\ and\ \bibinfo {author} {\bibfnamefont {J.~G.}\ \bibnamefont
  {Kirkwood}},\ }\bibfield  {title} {\bibinfo {title} {The statistical
  mechanical theory of transport processes. iv. the equations of
  hydrodynamics},\ }\href@noop {} {\bibfield  {journal} {\bibinfo  {journal}
  {The Journal of chemical physics}\ }\textbf {\bibinfo {volume} {18}},\
  \bibinfo {pages} {817} (\bibinfo {year} {1950})}\BibitemShut {NoStop}%
\bibitem [{\citenamefont {Malins}\ \emph {et~al.}(2013)\citenamefont {Malins},
  \citenamefont {Williams}, \citenamefont {Eggers},\ and\ \citenamefont
  {Royall}}]{malins2013tcc}%
  \BibitemOpen
  \bibfield  {author} {\bibinfo {author} {\bibfnamefont {A.}~\bibnamefont
  {Malins}}, \bibinfo {author} {\bibfnamefont {S.~R.}\ \bibnamefont
  {Williams}}, \bibinfo {author} {\bibfnamefont {J.}~\bibnamefont {Eggers}},\
  and\ \bibinfo {author} {\bibfnamefont {C.~P.}\ \bibnamefont {Royall}},\
  }\bibfield  {title} {\bibinfo {title} {Identification of structure in
  condensed matter with the topological cluster classification},\ }\href@noop
  {} {\bibfield  {journal} {\bibinfo  {journal} {J. Chem. Phys.}\ }\textbf
  {\bibinfo {volume} {139}},\ \bibinfo {pages} {234506} (\bibinfo {year}
  {2013})}\BibitemShut {NoStop}%
\bibitem [{\citenamefont {Fielding}(2014)}]{Fielding2014}%
  \BibitemOpen
  \bibfield  {author} {\bibinfo {author} {\bibfnamefont {S.~M.}\ \bibnamefont
  {Fielding}},\ }\bibfield  {title} {\bibinfo {title} {Shear banding in soft
  glassy materials},\ }\href@noop {} {\bibfield  {journal} {\bibinfo  {journal}
  {Reports on Progress in Physics}\ }\textbf {\bibinfo {volume} {77}},\
  \bibinfo {pages} {102601} (\bibinfo {year} {2014})}\BibitemShut {NoStop}%
\bibitem [{\citenamefont {Nicolas}\ \emph {et~al.}(2018)\citenamefont
  {Nicolas}, \citenamefont {Ferrero}, \citenamefont {Martens},\ and\
  \citenamefont {Barrat}}]{nicolas2018deformation}%
  \BibitemOpen
  \bibfield  {author} {\bibinfo {author} {\bibfnamefont {A.}~\bibnamefont
  {Nicolas}}, \bibinfo {author} {\bibfnamefont {E.~E.}\ \bibnamefont
  {Ferrero}}, \bibinfo {author} {\bibfnamefont {K.}~\bibnamefont {Martens}},\
  and\ \bibinfo {author} {\bibfnamefont {J.-L.}\ \bibnamefont {Barrat}},\
  }\bibfield  {title} {\bibinfo {title} {Deformation and flow of amorphous
  solids: Insights from elastoplastic models},\ }\href@noop {} {\bibfield
  {journal} {\bibinfo  {journal} {Reviews of Modern Physics}\ }\textbf
  {\bibinfo {volume} {90}},\ \bibinfo {pages} {045006} (\bibinfo {year}
  {2018})}\BibitemShut {NoStop}%
\bibitem [{\citenamefont {Royall}\ and\ \citenamefont
  {Malins}(2012)}]{Royall2012quench}%
  \BibitemOpen
  \bibfield  {author} {\bibinfo {author} {\bibfnamefont {C.~P.}\ \bibnamefont
  {Royall}}\ and\ \bibinfo {author} {\bibfnamefont {A.}~\bibnamefont
  {Malins}},\ }\bibfield  {title} {\bibinfo {title} {The role of quench rate in
  colloidal gels},\ }\href@noop {} {\bibfield  {journal} {\bibinfo  {journal}
  {Faraday Disc.}\ }\textbf {\bibinfo {volume} {158}},\ \bibinfo {pages} {301}
  (\bibinfo {year} {2012})}\BibitemShut {NoStop}%
\bibitem [{\citenamefont {Klotsa}\ and\ \citenamefont
  {Jack}(2013)}]{KlostaJCP13}%
  \BibitemOpen
  \bibfield  {author} {\bibinfo {author} {\bibfnamefont {D.}~\bibnamefont
  {Klotsa}}\ and\ \bibinfo {author} {\bibfnamefont {R.~L.}\ \bibnamefont
  {Jack}},\ }\bibfield  {title} {\bibinfo {title} {{Controlling crystal
  self-assembly using a real-time feedback scheme}},\ }\href
  {https://doi.org/10.1063/1.4793527} {\bibfield  {journal} {\bibinfo
  {journal} {The Journal of Chemical Physics}\ }\textbf {\bibinfo {volume}
  {138}},\ \bibinfo {pages} {094502} (\bibinfo {year} {2013})}\BibitemShut
  {NoStop}%
\bibitem [{\citenamefont {Tang}\ \emph {et~al.}(2016)\citenamefont {Tang},
  \citenamefont {Rupp}, \citenamefont {Yang}, \citenamefont {Edwards},
  \citenamefont {Grover},\ and\ \citenamefont {Bevan}}]{Tang2016}%
  \BibitemOpen
  \bibfield  {author} {\bibinfo {author} {\bibfnamefont {X.}~\bibnamefont
  {Tang}}, \bibinfo {author} {\bibfnamefont {B.}~\bibnamefont {Rupp}}, \bibinfo
  {author} {\bibfnamefont {Y.}~\bibnamefont {Yang}}, \bibinfo {author}
  {\bibfnamefont {T.~D.}\ \bibnamefont {Edwards}}, \bibinfo {author}
  {\bibfnamefont {M.~A.}\ \bibnamefont {Grover}},\ and\ \bibinfo {author}
  {\bibfnamefont {M.~A.}\ \bibnamefont {Bevan}},\ }\bibfield  {title} {\bibinfo
  {title} {Optimal feedback controlled assembly of perfect crystals},\
  }\href@noop {} {\bibfield  {journal} {\bibinfo  {journal} {ACS Nano}\
  }\textbf {\bibinfo {volume} {10}},\ \bibinfo {pages} {6791} (\bibinfo {year}
  {2016})}\BibitemShut {NoStop}%
\bibitem [{\citenamefont {Steinhardt}\ \emph {et~al.}(1983)\citenamefont
  {Steinhardt}, \citenamefont {Nelson},\ and\ \citenamefont
  {Ronchetti}}]{SteinhardtPRB83}%
  \BibitemOpen
  \bibfield  {author} {\bibinfo {author} {\bibfnamefont {P.~J.}\ \bibnamefont
  {Steinhardt}}, \bibinfo {author} {\bibfnamefont {D.~R.}\ \bibnamefont
  {Nelson}},\ and\ \bibinfo {author} {\bibfnamefont {M.}~\bibnamefont
  {Ronchetti}},\ }\bibfield  {title} {\bibinfo {title} {Bond-orientational
  order in liquids and glasses},\ }\href
  {https://doi.org/10.1103/PhysRevB.28.784} {\bibfield  {journal} {\bibinfo
  {journal} {Phys. Rev. B}\ }\textbf {\bibinfo {volume} {28}},\ \bibinfo
  {pages} {784} (\bibinfo {year} {1983})}\BibitemShut {NoStop}%
\bibitem [{\citenamefont {Yang}\ \emph {et~al.}(2012)\citenamefont {Yang},
  \citenamefont {Wu},\ and\ \citenamefont {Li}}]{yang2012generalized}%
  \BibitemOpen
  \bibfield  {author} {\bibinfo {author} {\bibfnamefont {J.~Z.}\ \bibnamefont
  {Yang}}, \bibinfo {author} {\bibfnamefont {X.}~\bibnamefont {Wu}},\ and\
  \bibinfo {author} {\bibfnamefont {X.}~\bibnamefont {Li}},\ }\bibfield
  {title} {\bibinfo {title} {A generalized irving--kirkwood formula for the
  calculation of stress in molecular dynamics models},\ }\href@noop {}
  {\bibfield  {journal} {\bibinfo  {journal} {The Journal of chemical physics}\
  }\textbf {\bibinfo {volume} {137}},\ \bibinfo {pages} {134104} (\bibinfo
  {year} {2012})}\BibitemShut {NoStop}%
\bibitem [{\citenamefont {Smith}\ \emph {et~al.}(2017)\citenamefont {Smith},
  \citenamefont {Heyes},\ and\ \citenamefont {Dini}}]{smith2017towards}%
  \BibitemOpen
  \bibfield  {author} {\bibinfo {author} {\bibfnamefont {E.}~\bibnamefont
  {Smith}}, \bibinfo {author} {\bibfnamefont {D.}~\bibnamefont {Heyes}},\ and\
  \bibinfo {author} {\bibfnamefont {D.}~\bibnamefont {Dini}},\ }\bibfield
  {title} {\bibinfo {title} {Towards the irving-kirkwood limit of the
  mechanical stress tensor},\ }\href@noop {} {\bibfield  {journal} {\bibinfo
  {journal} {The Journal of chemical physics}\ }\textbf {\bibinfo {volume}
  {146}},\ \bibinfo {pages} {224109} (\bibinfo {year} {2017})}\BibitemShut
  {NoStop}%
\bibitem [{\citenamefont {Bhowmik}\ \emph {et~al.}(2022)\citenamefont
  {Bhowmik}, \citenamefont {Hentschel},\ and\ \citenamefont
  {Procaccia}}]{bhowmik2022creep}%
  \BibitemOpen
  \bibfield  {author} {\bibinfo {author} {\bibfnamefont {B.~P.}\ \bibnamefont
  {Bhowmik}}, \bibinfo {author} {\bibfnamefont {H.~G.~E.}\ \bibnamefont
  {Hentschel}},\ and\ \bibinfo {author} {\bibfnamefont {I.}~\bibnamefont
  {Procaccia}},\ }\bibfield  {title} {\bibinfo {title} {Creep failure of
  amorphous solids under tensile stress},\ }\href
  {https://doi.org/10.1103/PhysRevE.106.034906} {\bibfield  {journal} {\bibinfo
   {journal} {Phys. Rev. E}\ }\textbf {\bibinfo {volume} {106}},\ \bibinfo
  {pages} {034906} (\bibinfo {year} {2022})}\BibitemShut {NoStop}%
\bibitem [{\citenamefont {Gelb}\ and\ \citenamefont
  {Gubbins}(1999)}]{GubbinsLang99}%
  \BibitemOpen
  \bibfield  {author} {\bibinfo {author} {\bibfnamefont {L.~D.}\ \bibnamefont
  {Gelb}}\ and\ \bibinfo {author} {\bibfnamefont {K.~E.}\ \bibnamefont
  {Gubbins}},\ }\bibfield  {title} {\bibinfo {title} {Pore size distributions
  in porous glasses: A computer simulation study},\ }\href
  {https://doi.org/10.1021/la9808418} {\bibfield  {journal} {\bibinfo
  {journal} {Langmuir}\ }\textbf {\bibinfo {volume} {15}},\ \bibinfo {pages}
  {305} (\bibinfo {year} {1999})}\BibitemShut {NoStop}%
\end{thebibliography}%
\newpage
\clearpage

\onecolumngrid
\break

\newcounter{equationSM}
\newcounter{figureSM}
\newcounter{tableSM}
\stepcounter{equationSM}
\setcounter{equation}{0}
\setcounter{figure}{0}
\setcounter{table}{0}
\setcounter{section}{0}
\makeatletter

\renewcommand{\theequation}{S\arabic{equation}}
\renewcommand{\thefigure}{S\arabic{figure}}
\renewcommand{\thesection}{\Alph{section}} 
\renewcommand{\thesubsection}{\arabic{subsection}} 

\begin{center}
  {\bf\large Appendices}
\end{center}

\newcommand{\figSnaps}{2f(inset)}

We provide additional details on various aspects of methods and results
\vspace{-8pt}\par
\begin{itemize}
\setlength{\itemsep}{0pt}
 \item Simulation methods
 \item Metrics used for the mechanical and structural characterization of breaking events, 
 \item Algorithm for detection of strand breakage. 
 \item Additional data for event correlations.
\item Additional data for macroscopic rheological response
\item Dependence of gel structure on preparation history.
\end{itemize}

\subsection{Details of simulation method}

We consider a Morse potential $U$ that mimics the short-ranged depletion interaction. 
Particles with mass $m$, velocity $v_i$ evolve with Langevin dynamics in a
simulation box with periodic boundaries :
\begin{equation}
m\frac{dv_i}{dt}=-\nabla  U -\lambda v_i +\sqrt{2\lambda k_B T}\xi_i
\label{SIeq_lan}
\end{equation}
where $\lambda$ is the friction constant and $\xi$ is the white noise.  The velocity damping time is $\tau_d=m/\lambda$.

We non-dimensionalise the system using the mean particle diameter $\bar \ell$ and the natural time scale $\tau=(m\bar{\ell}^2/k_B T)^{1/2}$, and taking $k_B T$ as the energy scale.  Then the non-dimensionalised Morse potential is $U_0=U/(k_{\rm B}T)$, note that $\alpha_0$ is the corresponding non-dimensionalised range parameter. The non-dimensionalised friction is $\lambda_0 = \lambda \tau/m$; we take $\lambda_0=10$ to mimic overdamped dynamics.   The integration time step is $\Delta t=0.001 \tau$.  To avoid crystallization we consider a size polydisperse system. We have taken $7$-types of particles with diameters $[0.88,0.92,0.96,1.0,1.04,1.08,1.12]$ (measured relative to $\bar \ell$) and with relative concentrations $[0.0062,0.0606,0.2417,0.3829,0.2417,0.0606,0.0062]$ to mimic a Gaussian distribution of diameters.
For colloidal motion, a natural time scale is the Brownian time $\tau_B  = \bar\ell^2\lambda/(24 k_BT)$ which is the typical time for an overdamped free particle to diffuse its radius.  For the parameters chosen here $\tau_B = 0.417\tau$.

We apply a constant shear stress $\sigma$ in the $xy$ plane following \cite{VezirovSM2015,CabrioluSM19}, which uses Lees-Edwards boundary conditions with flow along the $x$-direction.  The corresponding non-dimensionalized stress is $\sigma_0 = \sigma \bar\ell^3/(k_B T)$.  
This is maintained through a feedback control scheme that is implemented through the evolution of the shear rate 
\begin{equation}
\partial_t 
\dot{\gamma}=B[\sigma_0-\sigma_{xy}(t)]
\label{SIeq_rate}
\end{equation}
where $\sigma_{xy}$ is the total (non-dimensional) shear stress [measured from the virial] and $B$ is the damping parameter determining how quickly internal stress relaxes to imposed value. The system evolves by integrating the equations of motion Eqs.~\ref{SIeq_lan} and~\ref{SIeq_rate}, simultaneously to calculate the particle velocities and the strain rate, and applying a strain $\dot\gamma \Delta t$ in each time step.

\begin{figure}
\centerline{
\includegraphics[width=0.4\linewidth]{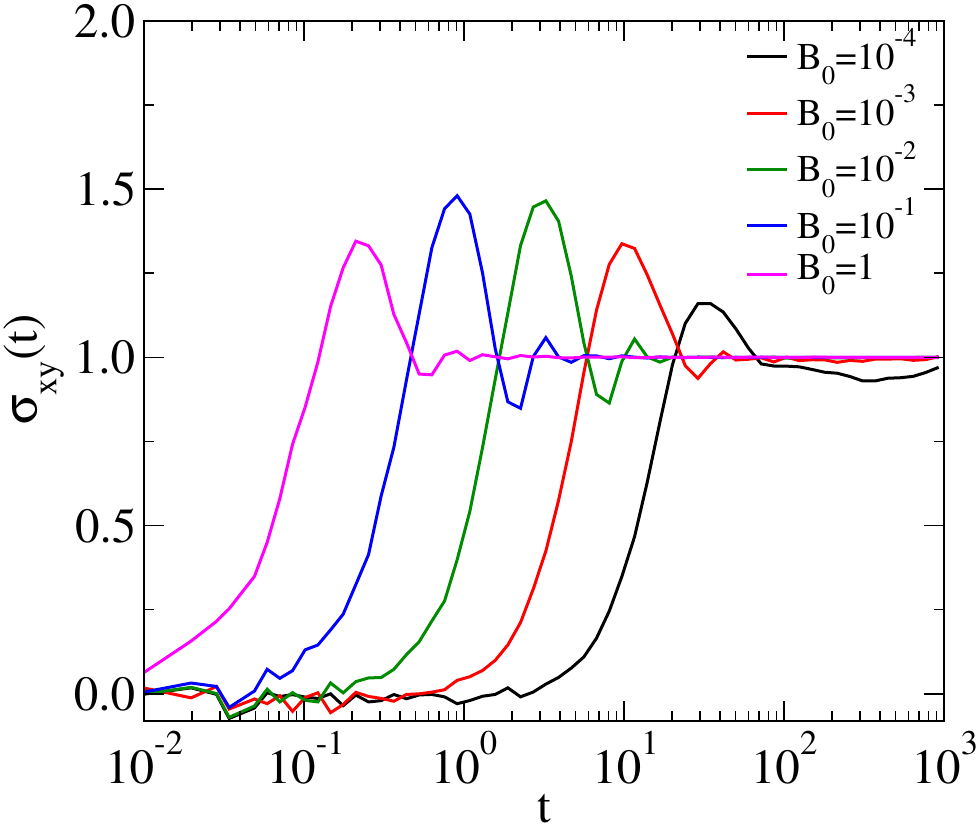}
}
\caption{\label{SI_fig_sigB} Virial stress (xy-component) as a function of time for different damping parameter $B_0$ while a gel ($\eprep=\eps_0=10$) is subjected to the constant stress shear protocol with $\sigma_0=1$. For small $B_0$, the system takes a long time to relax to the desired stress $\sigma_0$.}
\end{figure}

The dimensionless parameter $B_0=B\tau^2$ determines how fast the shear rate responds to changes in stress.  As discussed in~\cite{VezirovSM2015,CabrioluSM19}, this should be fast enough that the stress remains constant on the slow time scales associated with creeping and yielding.  We take $B_0=0.01$ which is large enough for our purposes, because the creeping flow is slow.  Fig.~\ref{SI_fig_sigB} shows the stress evolution starting from a quiescent state, showing that $\sigma_{xy}$ converges to $\sigma_0$ on a time of order $1$.  (Increasing $B$ makes the convergence faster, as expected.)  The overshoot in $\sigma_{xy}$ is due to the second order dynamics \eqref{SIeq_rate}, this effect also appears in the short-time oscillations of $\gamma$ in Fig. 1(a) of main text.

\subsection{Definitions of structural metrics}

To characterize strand-breaking events we compute two-fold bond orientational order parameter $q_2$~\cite{SteinhardtPRB83}, non-affine displacement $D^2_{\rm min}$~\cite{FalkPRE98D2min}, stress anisotropy $J_2$ of local stress computed by Irving and Kirkwood's method~\cite{irving1950statistical,yang2012generalized,smith2017towards,KrisSMNeck23}. We also perform topological cluster classification~\cite{malins2013tcc} to extract the structural information. All these quantities are single-particle measurements, as we now explain.

\paragraph{Bond orientational order parameter}
The bond orientational order parameter $q_l$ is calculated following~\cite{SteinhardtPRB83}:
\begin{align}
    q_{l}(i)=\sqrt{\frac{4\pi}{2l+1}\sum_{m=-l}^{l}\left|q_{lm}(i)\right|^2},\\
    q_{lm}(i)=\frac{1}{N_b(i)}\sum_{j=1}^{N_b(i)}Y_{lm}(\mathbf{r}_{ij}),
\end{align}
where $Y_{lm}$ are spherical harmonics and $N_b(i)$ is the number of neighbours of reference particle $i$. We consider the case $l=2$ when the quantity $q_2$ measures the stretching of a bond between two particles. Particles within the interaction range of particle $i$ are defined as its neighbour, and such definition is maintained for all the computations.

\paragraph{Non-affine displacement}
We consider non-affine displacement $D^2_{\rm min}$ per particle to quantify plasticity in the system. The non-affine displacement of a particle $i$ from time $t$ to $t+\Delta t$ is defined as~\cite{FalkPRE98D2min}:
\begin{equation}
    D^2_{\rm min}(i)(t,\Delta t)=\sum_j\sum_\alpha\left(r^\alpha_j(t)-r_i^\alpha(t)-\sum_\beta(\delta_{\alpha\beta}+\varepsilon_{\alpha\beta})\times\left[r_j^\beta(t-\Delta t)-r_i^\beta(t-\Delta t) \right] \right) ^2
\end{equation}
where the indices $\alpha$ and $\beta$ stand for the spatial coordinates and the index $j$ runs over $N_b(i)$ neighbours of the reference particle $i$. The local strain $\varepsilon_{\alpha\beta}$, which minimizes $D^2_{\rm min}$, is calculated as 
\begin{align}
    X_{\alpha\beta}&=\sum_j\left[r^\alpha_j(t)-r^\alpha_i(t)\right]\times\left[r^\beta_j(t-\Delta t)-r^\beta_i(t-\Delta t)\right],\nonumber\\
    Y_{\alpha\beta}&=\sum_j\left[r^\alpha_j(t-\Delta t)-r^\alpha_i(t-\Delta t)\right]\times  \left[r^\beta_j(t-\Delta t)-r^\beta_i(t-\Delta t)\right],\nonumber\\
    \varepsilon_{\alpha\beta}&=\sum_\gamma X_{\alpha\gamma} Y_{\gamma\beta}^{-1} - \delta_{\alpha\beta}.
\end{align}
We choose $\Delta t=1$. 

\paragraph{Stress anisotropy}
We measure local stress using a volume-averaged representation of the Irving–Kirkwood (IK) stress~\cite{irving1950statistical,yang2012generalized,smith2017towards,KrisSMNeck23}. Denoting $p_i$ as the momentum of particle $i$, $r_{ij}$ is the vector connecting particles $i$ and $j$, and $f_{ij}$ the corresponding inter-particle force, for a spatial region with volume $|\Omega|$, the $\mu \nu$ component of the IK stress is defined as 
\begin{equation}
    \sigma^{\mu \nu}_{|\Omega|}=-\frac{1}{|\Omega|}\left [ \sum_{i=1}^{N}\frac{1}{m_i}p_i^\mu p_i^\nu \theta_{i,\Omega}+\frac{1}{2}\sum_i^N \sum_{j\ne i}^N r_{ij}^{\mu }f_i^{\nu}\phi_{ij,\Omega}
       \right ]
\end{equation}
where $\theta_{i,\Omega}=1$ if particle $i$ is in $\Omega$ and zero otherwise; similarly, $\phi_{ij,\Omega}$ is the fraction of the straight line connecting particles $i,j$ that lies within $\Omega$. Taking $\Omega$ to be the entire simulation box $L^3$ gives the total stress $\sigma$, which can also be computed from the virial. In practice to measure the local stress $\sigma_\Omega(r)$ at point $r$, we take $\Omega$ as a small cube of side $l_{IK}$, centered at $r$. We choose $l_{IK} = 1.5$. To eliminate the noisy behaviour due to thermal fluctuations we average the stress value over a period of $20 \tau$~\cite{KrisSMNeck23} and then do the subsequent measurements.

We consider the second invariant $J_{2,\Omega}$ of the stress tensor to measure the anisotropy of the stress in local volume $\Omega$ as
\begin{equation}
    J_{2,\Omega}=\frac{1}{2}\mathrm{tr}\left( \left [ \sigma_\Omega-\frac{1}{3}\mathrm{tr}(\sigma_\Omega)\right ]^2 \right )
\end{equation}
Large $J_{2,\Omega}$ means there is more anisotropy in the volume $\Omega$ while a value zero means the volume $\Omega$ behaves like a simple fluid. We also note that since $J_{2,\Omega}$ is a scalar quantity, it is independent of
the orientation of the coordinate system. Particles in the volume $\Omega$ are assigned to have $J_{2,\Omega}$ value.

\paragraph{Topological cluster classification}
To investigate the structural change in the system we perform topological cluster classification. For each particle $i$, we compute (a) its number of neighbours $N_b$; (b) the number of fully-bonded tetrahedra in which it participates $n_{\rm tet}$ ; (c) the numbers of trigonal pyramids in which it participates, $n_{\rm tb}$. All the parameters of the TCC are kept the same as~\cite{malins2013tcc}.

\subsection{Strand breaking detection}
The gel is a complex topological object with load-bearing strands whose thickness and length are highly dependent on interaction strength and preparation history. Upon shear deformation, such a strand experiences force, becomes thin with the formation of the neck and ultimately breaks. In order to detect such a breaking event we rely on the topological measure namely the chemical distance $\ell_d$ between two neighboring particles~\cite{tsurusawa2019direct}. The chemical distance $\ell_d$ is the shortest path between particle $i$ and $j$ traversing through the neighbouring particles. If two particles are neighbour to each other, $\ell_d=1$, else $\ell_d>1$. 

When a strand breaks the pair of neighbouring particles at the breaking point will have a sudden jump $\Delta \ell_d$ in their chemical distance from $\ell_d=1$ to some arbitrarily higher value. We consider $\Delta \ell_d>6$ to identify the breaking event (see however below). For simplicity, we insist that particles that have already participated in breaking events cannot be involved in any later ones. 

\begin{figure}
\centerline{(a)
\includegraphics[width=0.16\linewidth]{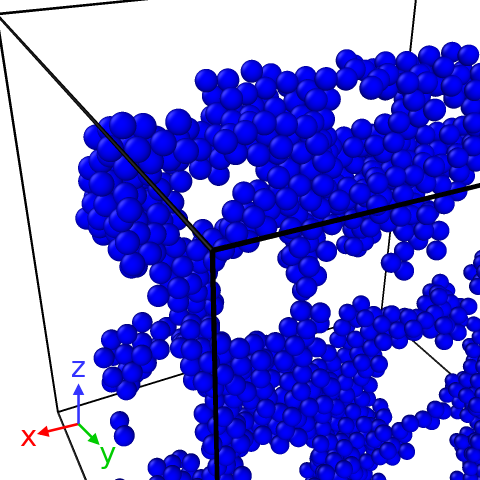}\qquad
(b)
\includegraphics[width=0.16\linewidth]{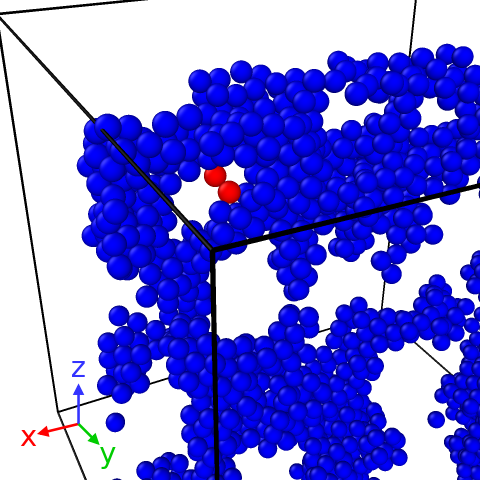}\qquad
(c)
\includegraphics[width=0.16\linewidth]{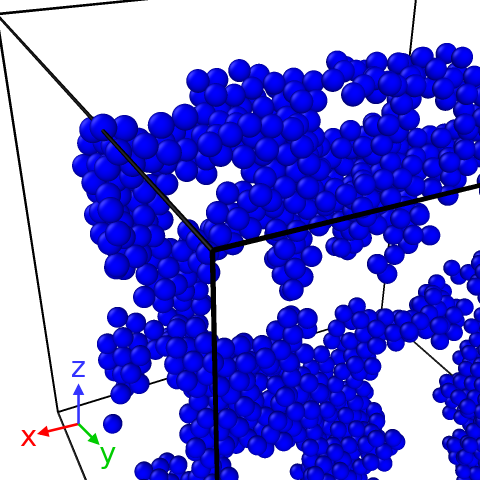}
}
\caption{\label{SI_fig1} A part of the system at a different time of simulation showing two arms that are away before [shown in (a)], come closer and particles from each arm (in red) become neighbouring pair [shown in (b)], and a later time those two particles move apart [shown in (c)].  Even though $\Delta \ell_d>6$ at the breaking time, our algorithm does not detect this as a breaking event.}
\end{figure}

We note that due to the shear deformation, two strands may come closer and particles from each strand may become neighbors to each other temporarily and detach after some time as shown in Fig.~\ref{SI_fig1}. Such transient formation of neighboring pairs can also happen due to the diffusion of the particles due to thermal fluctuation. This can lead to false detection of strand-breaking events according to the criteria $\Delta \ell_d >6$.  To avoid such artifacts, we only identify a strand-breaking event if $\Delta \ell_d>6$ at the breaking time $t_b$ and that $\ell_d=1$ for $t_b-\delta t < t < t_b$ and that $\ell_d>1$ for $t_b < t < t_b+\delta t$.  We take $\delta t = 20\tau$, results depend weakly on this parameter.

\subsection{Further data for strand breaking events}

\begin{figure}
\flushleft \hspace{2cm}(a) \hspace{3.2cm} (b) \hspace{3.4cm} (c) \hspace{2.9cm} (d)
\centerline{
\includegraphics[width=0.22\linewidth]{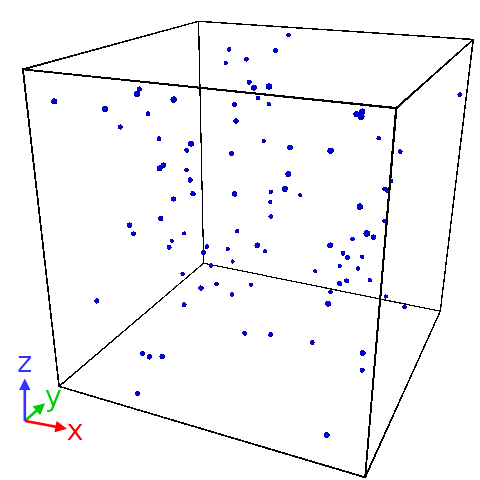}
\includegraphics[width=0.22\linewidth]{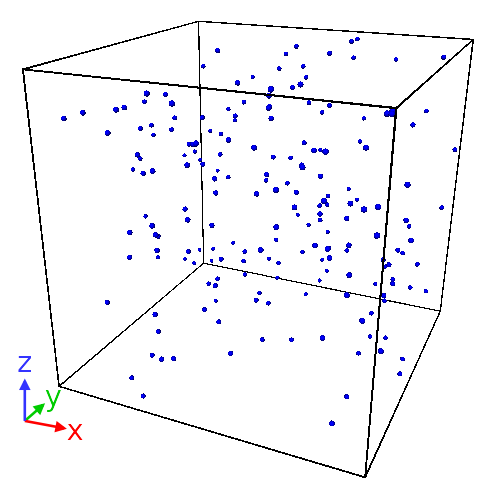}
\includegraphics[width=0.22\linewidth]{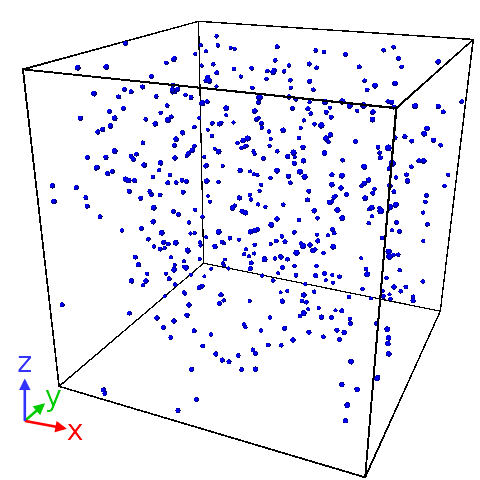}
\includegraphics[width=0.26\linewidth]{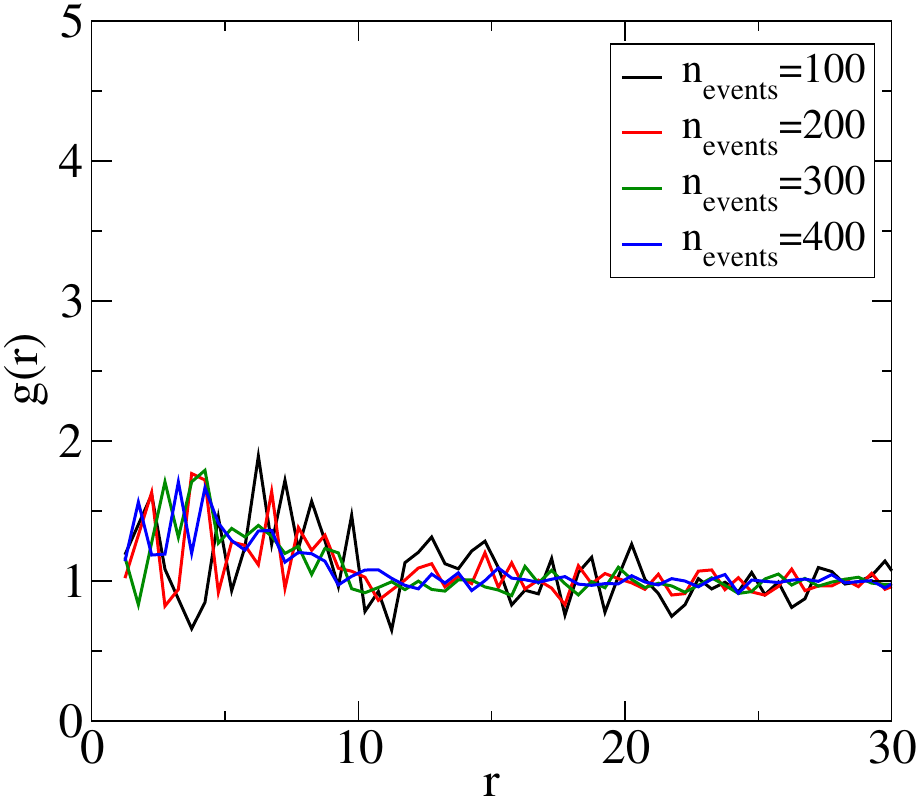}}
\caption{\label{SI_fig_gr} (a,b,c) Snapshots of locations of the strand-breaking events after $n_{\rm events}=100,200,400$, respectively.  [See also Fig.~\figSnaps\ of main text.] (d) Radial distribution function of the location of strand breaking events after different number of $n_{\rm events}$.} 
\end{figure}

\paragraph{Spatially uncorrelated failure events}
In the main text, we show that there is no spatial correlation in the location of the strand-breaking events. In Figs.~\ref{SI_fig_gr}(a-c) we show the snapshot of the location of the breaking events at different times. We see that the points are scattered in space and apparently there is no spatial correlation between them. To quantify that here we compute the radial distribution function of the breaking points after some number of events $n_{\rm events}$. In Fig.~\ref{SI_fig_gr}(d) we show the radial distribution of the location of events for different $n_{\rm events}$. The distribution exhibits ideal gas behaviour for the whole range of $r$, inferring the uncorrelated nature of event location. 

\begin{figure}
\centerline{
\includegraphics[width=1\linewidth]{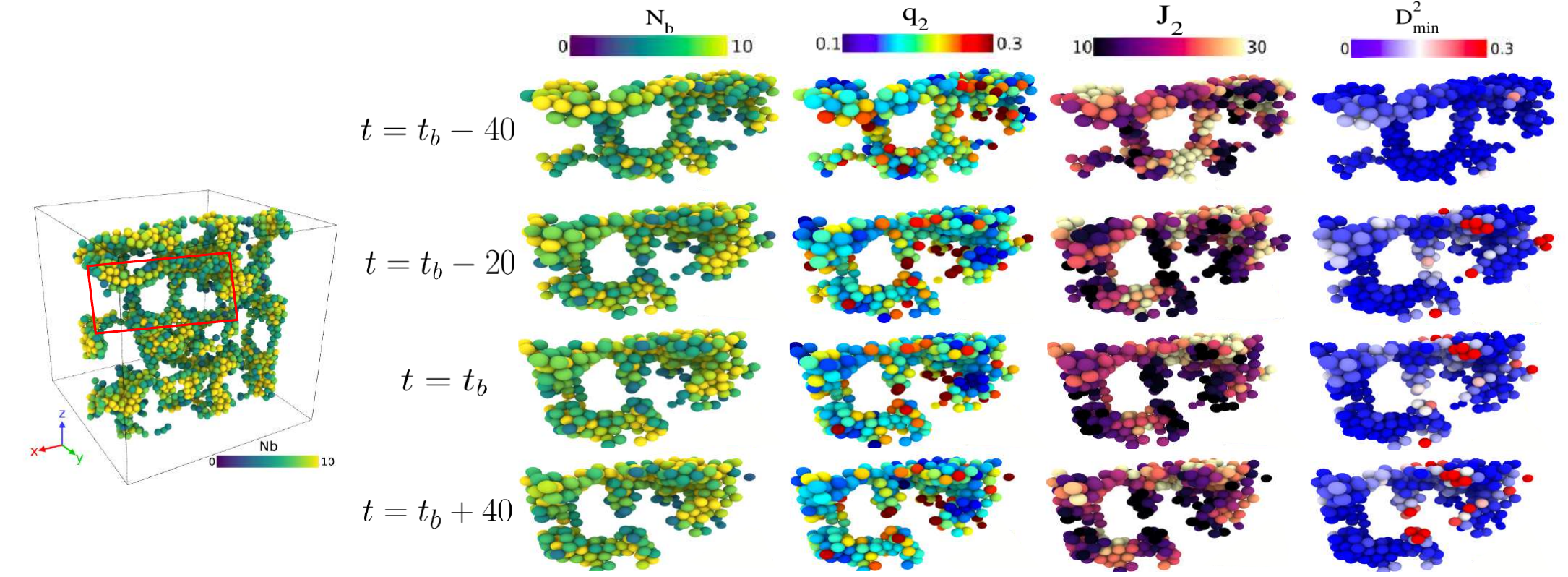}
}
\caption{\label{SI_fig_strand} (left) Rendering of a slice through a gel, to visualize the strands for $\eps_0=10$. The highlighted boxed area denotes the particular strand whose evolution we will monitor around the breaking time $t_b$. Particles are colored according to the value of different metrics as, the first column is for the number of neighbour, the second is for two-fold bond orientation parameter $q_2$, the third column is with stress anisotropy $J_2$, and the fourth column is for non-affine displacement $D^2_{\rm min}$. }
\end{figure}

\paragraph{Structural change around strand breaking}
In the main text, we show how different quantities vary with time when a strand going to break. Here we provide a pictorial representation of that. In Fig.~\ref{SI_fig_strand} we show the variation of different quantities as a measure of structural and mechanical change during the strand-breaking event. The strand is very thick long before breaking at $t_b-40$; As time elapses the quantities change interestingly. The location of the strand where it is going to break becoming thinner (smaller $N_b$), more stretched (larger $q_2$), and more liquid like behaviour exhibited by smaller stress anisotropy. A  larger $D^2_{\rm min}$ value around the location indicates the region undergoes more plastic displacement while it breaks.

\subsection{Further data for macroscopic rheology}
\begin{figure}
\centerline{
\includegraphics[width=0.249\linewidth]{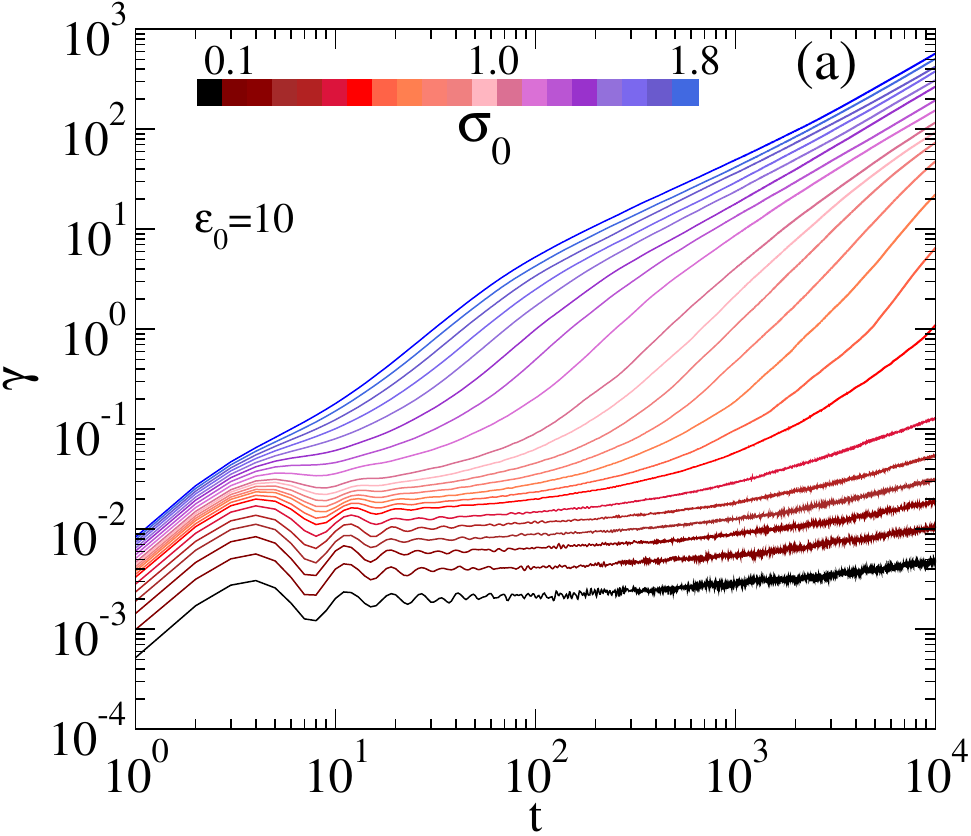}
\includegraphics[width=0.253\linewidth]{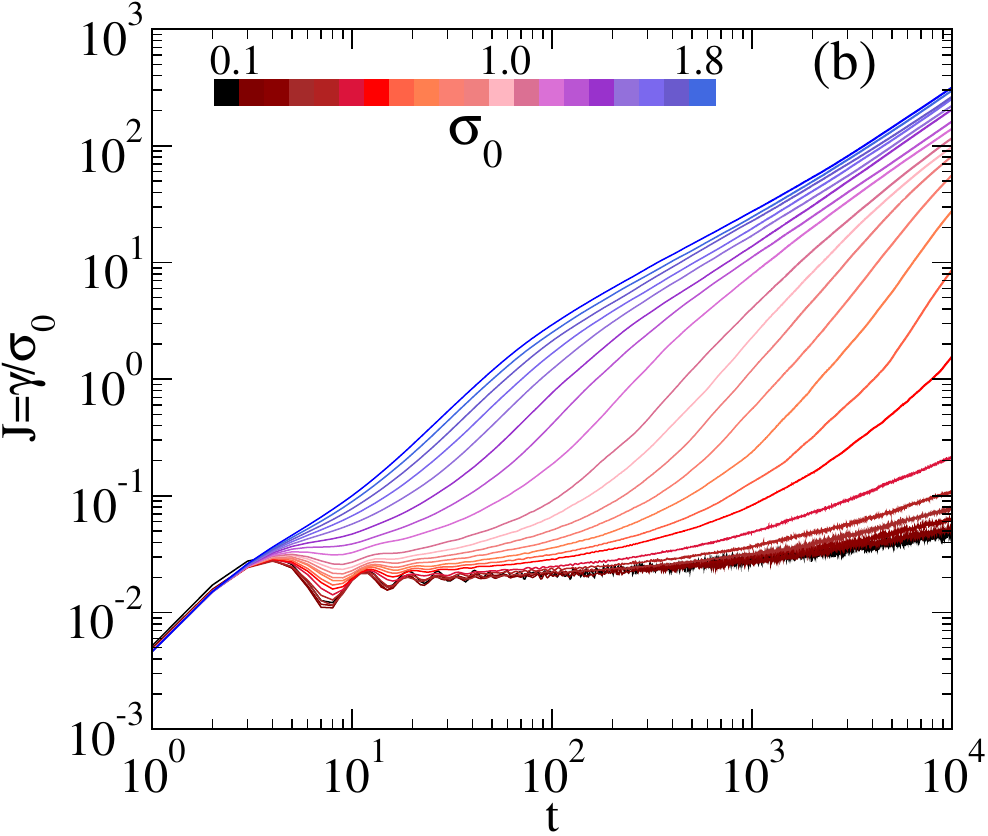}
\includegraphics[width=0.246\linewidth]{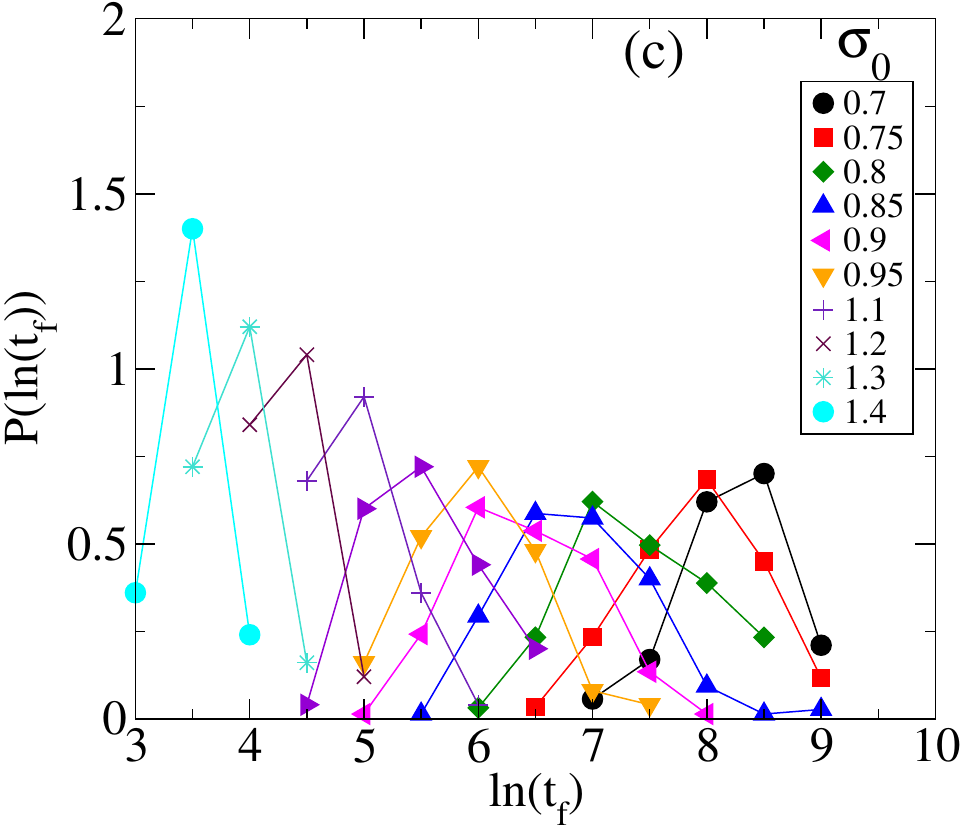}
\includegraphics[width=0.238\linewidth]{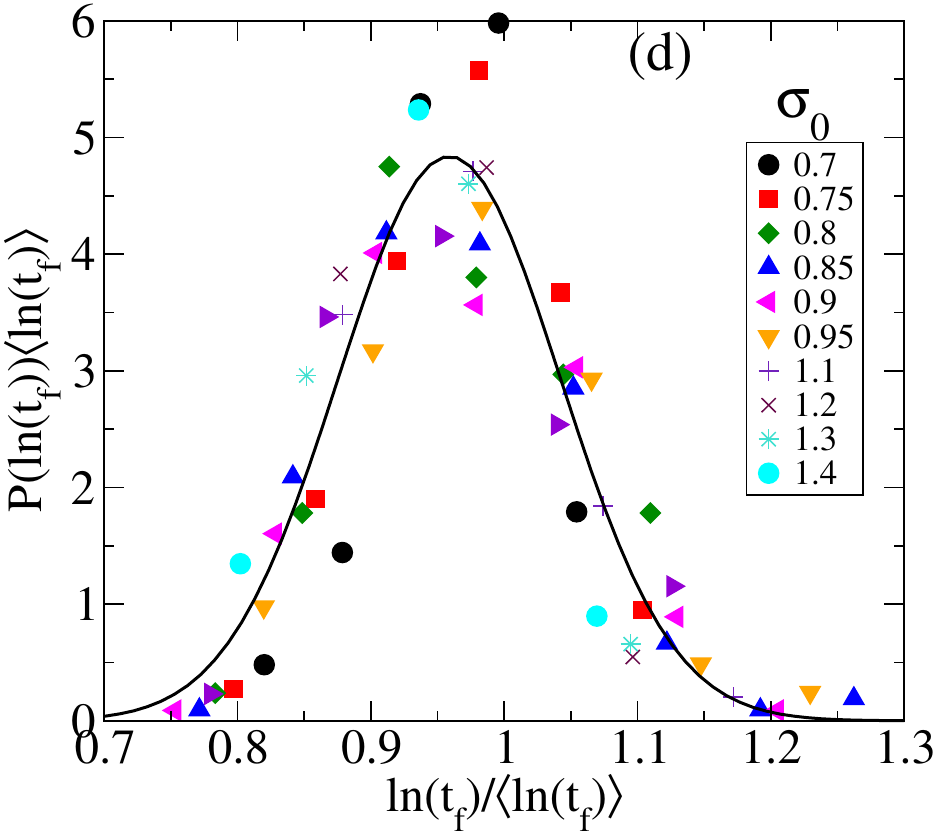}
}
\caption{\label{SI_fig_Jdisttf} (a) Strain vs time and (b) compliance $J=\gamma/\sigma_0$ vs time, for various imposed stresses  $\sigma_0$, at  $\eps_0=10$. (c) Distribution of failure time obtained from different samples for different $\sigma_0$. (d) Scaled distribution showing data collapse. The solid line a Gaussian fit. }
\end{figure}

To support Figs.4 (a,b) of the main text, we show in Figs.~\ref{SI_fig_Jdisttf}(a,b) the average strain and compliance of gels with $\eps_0=\eprep=10$. These are the data from which we extract the effective viscosity $\sigma/\dot\gamma$, by finite time-differences. The normalized strain variance in Fig.4(b) of the main text is computed as
\begin{equation}
(\Delta\gamma)^2 = \frac{\langle \gamma^2(t) \rangle - \langle \gamma(t) \rangle^2}{ \langle \gamma(t) \rangle^2}
\end{equation}
which is then plotted parameterically as a function of $\gamma = \langle \gamma(t)\rangle$.  The data collapse for small $\sigma_0$ in Fig.~\ref{SI_fig_Jdisttf}(b) shows that a linear-response regime exists for the solid-like response to shear (for larger $\eps_0$).

As discussed in the main text, $(\Delta\gamma)^2$ is consistently maximal for $\gamma\approx \gamma_c=0.4$ which we identify as a critical strain.  For each sample $\alpha$, we define the individual failure time as $t_{\rm f}^{(\alpha)}$ as the time at which $\gamma=\gamma_c$.  Averaging over samples gives the average failure time $t_{\rm f}$ discussed in the main text. In  Fig.~\ref{SI_fig_Jdisttf}(c,d) we show the distributions of these times.  As expected the peak of the distribution decreases with increasing $\sigma_0$ while keeping their Gaussian form.  While rescaled with their mean, we find a nice collapse of data as shown in Fig.~\ref{SI_fig_Jdisttf} (d). It should be noted that a similar scaling form has also been observed for the creeping of amorphous solid under tensile force~\cite{bhowmik2022creep}.

\subsection{Further data for dependence on $\eprep$ and $t_{\rm w}$}

\begin{figure}[t]
\centerline{
\includegraphics[width=1\linewidth]{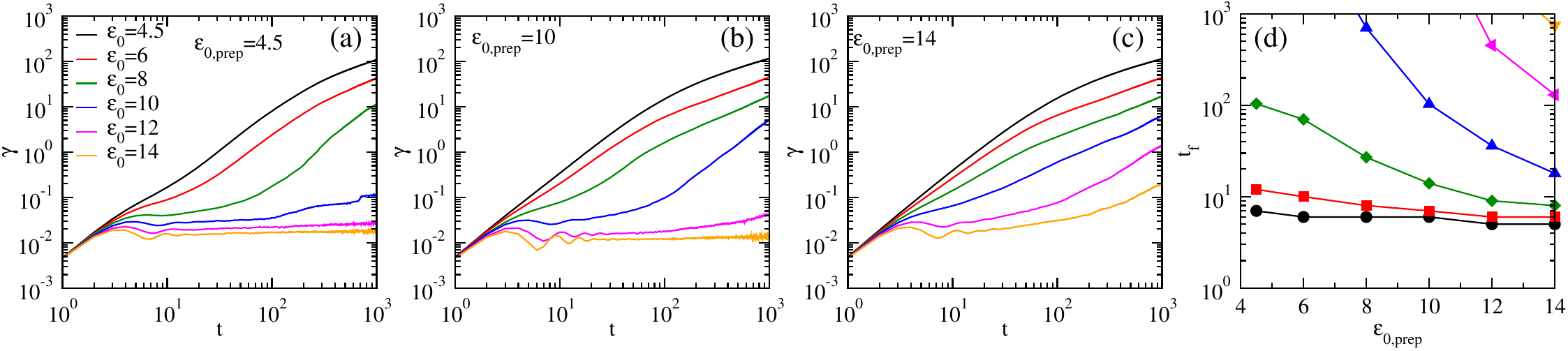}
}
\caption{\label{SI_fig_eprep} Strain against time for different $\eps_0$ during shear for a gel prepared with (a) $\eprep=4.5$, (b) $\eprep=10$, and (c) $\eprep=14$. (d) Failure time $t_{\rm f}$ as a function of $\eprep$ for several $\eps_0$. Legends in (a) is common to all panels.}
\end{figure}

Recall from Fig. 5(b) of main text  that the failure time $t_{\rm f}$ depends significantly on the preparation protocol of the gel.
Fig~\ref{SI_fig_eprep} explores these effects in more detail.
We show the strain evolution for three different gel prepared with different $\eprep$ in Figs.~\ref{SI_fig_eprep}(a,b,c). However, to reduce numerical computation we simulate up to $t=10^3$ and relax the condition of the definition of failure time by taking $t_{\rm f}$ as the time to reach $\gamma=0.1$. In Fig.~\ref{SI_fig_eprep}(d) we show $t_{\rm f}$ as a function of  $\eps_0$ and $\eprep$. 

\begin{figure}[h]
\centerline{
\includegraphics[width=.96\linewidth]{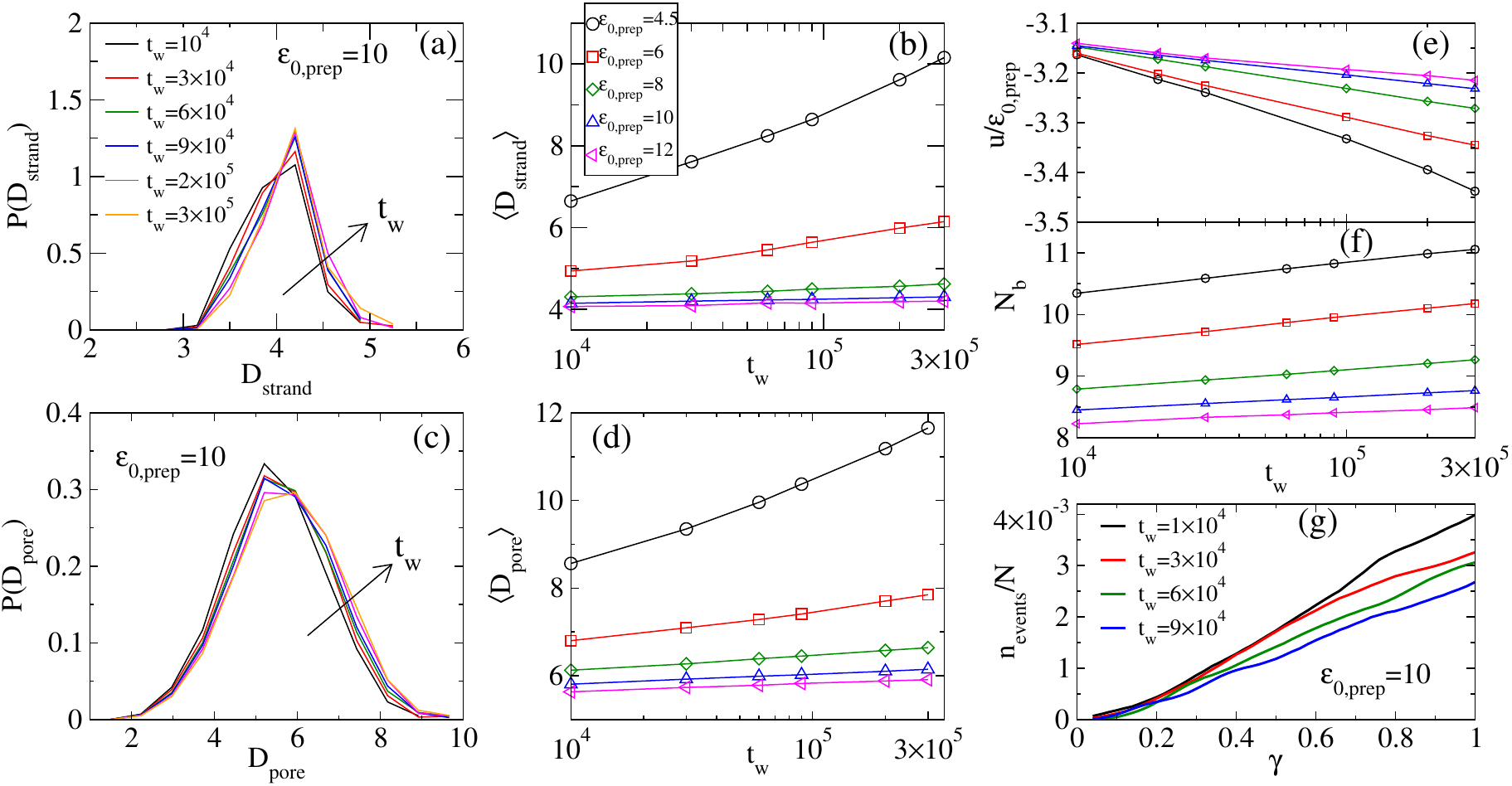}}
\caption{\label{SI_fig_struc} Effect of $t_{\rm w}$: Distribution of (a) strand thickness and (c) pore size for a gel with $\eprep=10$ for different waiting time $t_{\rm w}$. (b)Average strand thickness and (d) average pore diameter as a function of $t_{\rm w}$ for different $\eprep$. (e) Per particle potential energy and (f) the average number of neighbours as a function of $t_{\rm w}$ for different $\eprep$. (g) Number of breaking events as a function of accumulated strain for different $t_{\rm w}$ for a gel with $\eprep=\eps_0=10$. [(a) shares color coding with (c); (b),(d),(e), and (f) share the same color code]}
\end{figure}

In addition, Fig.~\ref{SI_fig_struc} shows how the gel structure depends on $t_{\rm w}$, for different $\eprep$.  (These results are obtained at $t=0$, before any shear stress is applied.) Following~\cite{GubbinsLang99}, we measure the pore size of the gel and extend the similar idea to measure the strand thickness. To measure the pore size, we find the largest possible sphere that can fit in the void space without overlapping with any colloidal particles. Similarly, to measure the strand thickness, we find the largest possible sphere that can fit in the colloidal space without encompassing the void space. The distribution of pore diameters and strand thickness measured within the gel has a strong dependence on $t_{\rm w}$, as shown in Figs.~\ref{SI_fig_struc}(a) and~\ref{SI_fig_struc}(c), respectively. Their average, as shown in Fig.~\ref{SI_fig_struc}(b) and (d), increases with  $t_{\rm w}$.  This manifests the coarsening of gel with a thicker arm. It should be noted that for a given waiting time, pore diameter and strand thickness also increase with decreasing interaction strength. 

Furthermore, we measure the per particle number of neighbours and energy for a gel prepared with different $\eprep$. As shown in Figs.~\ref{SI_fig_struc}(e) and~\ref{SI_fig_struc}(f), their variation with $t_{\rm w}$ again reveals the annealing behaviour with a more stable structure as $t_{\rm w}$ increases. 

Lastly, we measure the strand-breaking statistics. Gel structures sampled at different $t_{\rm w}$ are subjected to shear keeping all other parameters at fixed. The variation of $n_{\rm events}/N$ against accumulated strain is shown in Fig.~\ref{SI_fig_struc}(g). As expected gel with larger $t_{\rm w}$ experiences fewer strand breaking events than that with smaller $t_{\rm w}$. Altogether, this analysis showcases how interaction strength and waiting time influence the gel structure and hence its mechanical stability.

\end{document}